\documentclass[aps,prx,reprint,superscriptaddress,longbibliography,floatfix]{revtex4-2}

\usepackage{amsmath,amssymb}
\usepackage{graphicx}
\usepackage{booktabs}
\usepackage{xcolor}
\usepackage[colorlinks=true,allcolors=blue]{hyperref}

\graphicspath{{figs/}}

\newcommand{\pstar}{p^{*}}

\begin{document}

\title{Do Quantum AIs Dream in Paths? Path-Integral Slow Thinking through
Grover Interference}
%

\author{Xiansheng Cai}
\affiliation{Institute of Theoretical Physics, Chinese Academy of Sciences, Beijing 100190, China}

\author{Xiu-Hao Deng}
\affiliation{Shenzhen International Quantum Academy, Shenzhen 518048, China}
\affiliation{Shenzhen Branch, Hefei National Laboratory, Shenzhen 518048, China}

\author{Kun Chen}
\email{chenkun@itp.ac.cn}
\affiliation{Institute of Theoretical Physics, Chinese Academy of Sciences, Beijing 100190, China}

\date{\today}

\begin{abstract}
Reinforcement learning with verifiable rewards enables large language models
to think slowly through long chains of thought, but the same training can
induce policy collapse: probability concentrates onto a few successful
trajectories, and exploratory diversity erodes.  We ask whether quantum AI
can realize slow thinking differently.  We formulate slow thinking
as coherent dynamics over reasoning trajectories, a discrete path
integral in which action sequences coexist in superposition and recombine
before measurement.  In our trainable realization, an exact verifier
partitions the ensemble into collective accepted and rejected components
that interfere under Grover amplitude amplification; the working resource is
this ensemble-level interference.  A finite Grover evolution is maximized when the success
probability before amplification lies at an analytically determined value
below one, so inference itself defines an interior training target and
removes the monotonic pressure toward unit success.  In exact statevector
simulations of a $2\times3$ sliding puzzle with interdependent moves, Grover
training reaches accuracy $0.95$ on a $32$-question training set at one
round, against $0.73$ for the strongest classical control.  On held-out
questions, specialization has a cost: an untrained uniform policy read out
through the same amplification remains the strongest reference on this
solution-dense benchmark.  Quantum training preserves far more held-out
accuracy than classical training: at four rounds with matched
reasoning-circuit applications, the table-based and neural quantum models
reach $3.2$ and $3.9$ times the held-out accuracy of the strongest classical
controls.  For policies of fixed size trained separately at each
amplification budget, the number of supported training questions grows
faster with the budget than with matched classical repetition.  These
results establish a Grover-based realization of path-integral slow thinking,
in which the interior target set by inference preserves exploratory path
diversity and ensemble-level interference converts interior success
probabilities into verified performance.
\end{abstract}

\maketitle

\section{Introduction}

Recent advances in large language model
(LLM) reasoning have been driven in part by \emph{chain-of-thought}
generation~\cite{wei2022chain}.  Rather than producing a final answer directly,
an LLM generates a sequence of intermediate tokens, with each new token
conditioned on the problem and the preceding sequence.  A complete reasoning
sequence can therefore be viewed as a stochastic trajectory through possible
intermediate steps.  Because each step admits multiple continuations,
increasing the sequence length expands both the depth of the computation and
the space of possible reasoning trajectories.  This enlarged space allows the
model to compose more intermediate operations and thereby reach solutions that
are inaccessible to shorter trajectories.  Increasing the reasoning horizon at
inference---known as \emph{test-time scaling}---can therefore enable a model to
solve harder problems without increasing its number of
parameters~\cite{deepseek2025r1,snell2024scaling,brown2024monkeys}.  Continued progress in
coherent quantum information processing raises a natural question: could a
quantum AI model acquire an analogous form of slow
thinking~\cite{kahneman2011thinking} through quantum
dynamical evolution?  If so, how would its reasoning differ fundamentally
from classical slow thinking, how could such a model be trained, and what
computational or learning advantages might coherence provide?

The first question has a natural dynamical answer.  An LLM defines a
statistical distribution over reasoning trajectories, from which each
inference run samples one realized trajectory.  A quantum model instead
propagates amplitudes coherently through multiple reasoning histories and
combines them by a path integral before the final measurement.  Let $x$ be the
input problem, $y$ a possible final answer, and
$\gamma=(z_1,\ldots,z_T)$ a complete sequence of $T$ intermediate reasoning
states generated from $x$, with $z_t$ denoting the state at step $t$.  We write
$\gamma:x\to y$ when this trajectory contributes
to the final answer $y$.  Schematically,
\begin{equation}
\begin{aligned}
P_{\mathrm{LLM}}(y|x)
&\propto \sum_{\gamma:x\to y}
e^{-I_{\mathrm{LLM}}[\gamma|x]},\\
P_{\mathrm Q}(y|x)
&\propto \left|\sum_{\gamma:x\to y}
A_{\mathrm Q}[\gamma|x]\right|^2 .
\end{aligned}
\label{eq:reasoning-path-integral}
\end{equation}
Here $I_{\mathrm{LLM}}[\gamma|x]$ is the surprisal of a reasoning trajectory,
whereas
$A_{\mathrm Q}[\gamma|x]=|A_{\mathrm Q}[\gamma|x]|
e^{iS_{\mathrm Q}[\gamma|x]/\hbar}$ is the corresponding quantum trajectory
amplitude.  The effective phase action $S_{\mathrm Q}$ is defined modulo
$2\pi\hbar$, and $\hbar$ fixes the unit in which the phase is measured and may
be set to one.  The
proportionality signs suppress normalization and readout-dependent factors.
The first expression is a
statistical sum: an individual run realizes one trajectory.  The second is a
path integral over complex amplitudes, in which multiple histories coexist in
superposition and can interfere before measurement.  Increasing the
reasoning horizon enlarges the space of histories in both cases, but only the
quantum dynamics can recombine their amplitudes coherently.  This distinction
suggests that a longer reasoning horizon could enlarge the available
reasoning space while interference selects the final answer without first
collapsing onto one trajectory.  This is not yet an advantage: coherence alone
neither selects useful interference nor explains how a model should preserve
and exploit a large space of reasoning trajectories.  To formulate this
problem, we first examine how classical systems convert additional inference
time into a search over possible trajectories.

AlphaGo provides a canonical state-based realization of this idea.  Here a
\emph{policy} is a parameterized rule that assigns probabilities to available
actions, whereas a value network estimates the expected future outcome of a
state.  AlphaGo combined these learned networks with Monte Carlo tree search
(MCTS)~\cite{silver2016alphago,browne2012mcts}.  Because Go supplies explicit
board states, legal actions, and known transitions, MCTS can repeatedly expand
a persistent tree, evaluate possible continuations, and propagate their
outcomes backward.  This construction separates learning from slow thinking:
reinforcement learning improves the policy and value networks that guide the
search, whereas MCTS converts additional inference-time computation into
deeper and broader exploration without changing the network parameters.
Reasoning LLMs pursue the same objective under a different constraint.  Their
effective state is the complete token history rather than a compact,
task-defined state with a known transition graph, making persistent tree
construction and state reuse substantially more difficult.  Modern reasoning
models therefore commonly use trajectory-level \emph{reinforcement learning
with verifiable rewards} (RLVR): complete reasoning trajectories are sampled,
a verifier scores their final answers, and sequence probabilities are updated
without constructing a persistent tree of explicit intermediate states.
DeepSeekMath introduced group relative policy optimization (GRPO), a method
that estimates each trajectory's relative advantage from other trajectories
sampled for the same problem, without a separately trained value model (often
called a critic)~\cite{shao2024deepseekmath}.  DeepSeek-R1 showed that
GRPO-based RLVR
can produce long reasoning sequences, self-correction, and strategy changes
without labelled intermediate steps~\cite{deepseek2025r1}.

This trajectory-level training is powerful, but it has an intrinsic tendency to
improve single-attempt accuracy at the expense of exploratory diversity.
Suppose that a model initially possesses two distinct strategies for solving
the same problem, each with modest success probability.  Early reward
fluctuations may favor one strategy.  Reinforcement then increases its
probability, causing it to be sampled and rewarded still more frequently,
while the competing strategy gradually disappears.  The resulting model may
achieve high single-attempt accuracy by exploiting the dominant strategy, yet
lose the diversity required to explore alternative solutions.  Additional
test-time computation then produces variations of essentially the same
reasoning path rather than opening new ones~\cite{yue2025rlvr,
yuan2026overtraining,cui2025entropy}.  This phenomenon is known as
\emph{policy collapse}.  MCTS controls an analogous loss of exploration through
an explicit exploration term: branches with high estimated value are favored,
but less-visited branches continue to be sampled.  Trajectory-level RLVR has no
direct analogue of this node-wise visitation information because it updates a
global distribution over complete sequences.  Entropy regularization and
confidence constraints can slow the collapse~\cite{cui2025entropy,chen2025confidence}, but no
general mechanism currently combines reliable outcome optimization with
sustained diversity of reasoning trajectories.  This leaves the training
question posed above: can a coherent inference map optimize verified outcomes
without erasing the path diversity on which it acts?

\begin{figure*}[!t]
  \includegraphics[width=\textwidth]{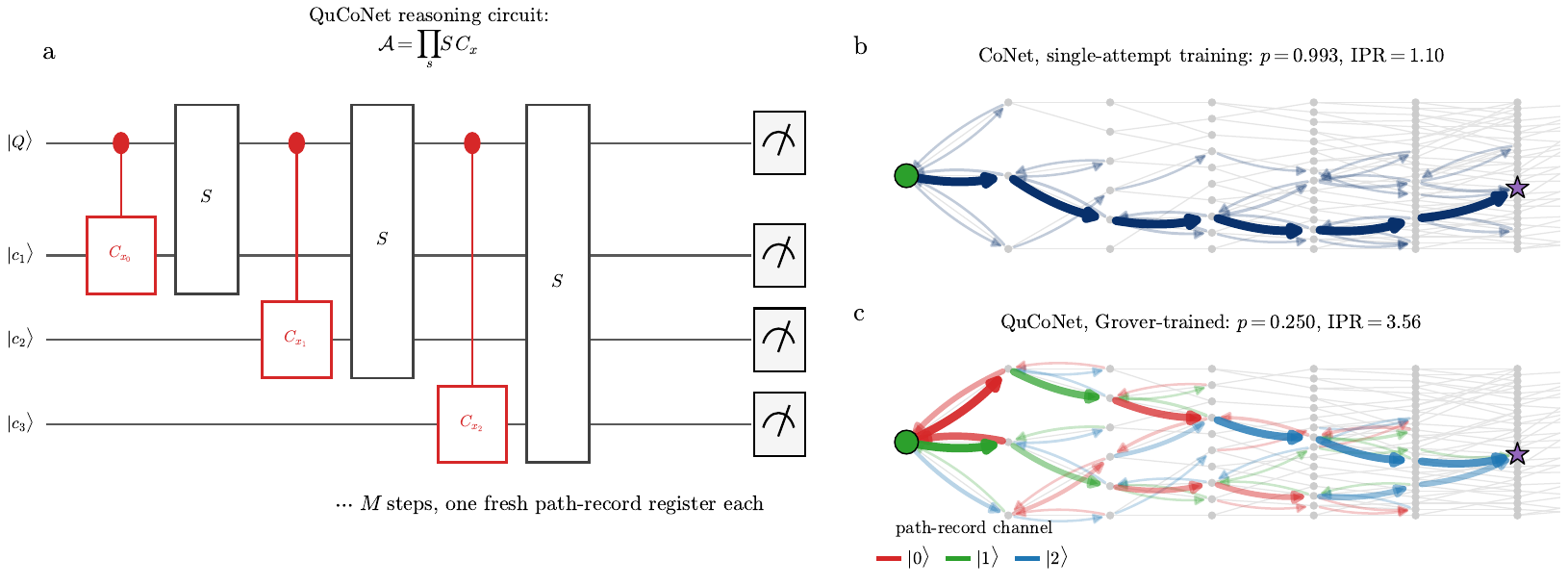}
  \caption{\textbf{Coherent inference preserves multiple reasoning trajectories.}
  (a)~The table-based quantum concept-network model (QuCoNet) records each move
  in a fresh auxiliary register, called a which-path register.  Measuring these
  records reduces the reasoning circuit to the corresponding
  classical Markov chain,
  whose probabilities are computed exactly; retaining the records coherently prepares all
  trajectories for Grover amplification in Fig.~\ref{fig:collapse}(a)
  (Supplemental Material).  (b),(c)~Routing for one training question
  (training-set size $B=32$, question-pool seed~$1$; start, green circle;
  target, purple star).  Nodes are ordered by distance from the start, and edge
  width gives the success-probability flow through that edge.
  Single-attempt training concentrates the
  classical walker on one route.  Grover-based training retains a branching
  ensemble of paths distinguished by orthogonal which-path records at the
  interior target.  The inverse participation ratio (IPR) is the effective number of
  paths; panel (c) shows a favorable case, while the accuracy-weighted mean IPR
  is $1.6$ at $q=3$, where $q$ counts reasoning-circuit applications.  The panels show path weights,
  not interference between the displayed paths.}
  \label{fig:paradigm}
\end{figure*}

Existing quantum AI provides important ingredients for this question, but it
does not yet realize slow thinking as test-time scaling of a learned reasoning
model.  Quantum projective-simulation agents use quantum walks to accelerate
deliberation over an episodic-memory
network~\cite{paparo2014agents,katabarwa2017hamiltonian,
franceschetto2024projective}; parametrized quantum circuits serve as
trainable policies over observed states~\cite{jerbi2021parametrized}; and
amplitude amplification accelerates inference in trained quantum generative
models~\cite{zeng2019generative} and reward discovery in episodic
reinforcement learning, in theory and in a photonic
experiment~\cite{hamann2022hybrid,saggio2021experimental}, where the policy
being learned remains classical.  Closest to our setting, quantum Markov
decision processes propagate coherent superpositions of multistep
state--action trajectories and amplify those with high
returns~\cite{su2025qmdp,su2025grover}.  These approaches provide important
ingredients, but leave unresolved the central question of quantum slow thinking:
can a fixed-size learned quantum model turn extra coherent inference computation
into greater verified problem-solving capability while avoiding the
policy-collapse trade-off of classical trajectory-level RLVR---the concentration
of probability on only a few successful reasoning trajectories?

In this Article, we formulate quantum slow thinking as a discrete path
integral over reasoning trajectories, realized as coherent dynamics in a
trainable quantum reasoning network.  We realize this framework with coined
quantum walks in two forms~\cite{aharonov1993quantum,ambainis2001onedim,kempe2003overview}.
The first, QuCoNet, is a table-based quantum extension of the concept-network
model (CoNet), an effective network description of multistep
reasoning~\cite{cai2025criticality,hu2025freezing}: each
problem state carries a directly optimized local unitary that distributes
amplitude among the allowed next actions.  The second is a neural quantum AI
model in which a parameter-sharing network generates this local unitary from
the question and the preceding action sequence.  In both models, a trainable
reasoning circuit $\mathcal A_\theta$, with $\theta$ denoting the model
parameters, prepares a coherent ensemble of action
sequences.  A verifier separates this ensemble into
collective accepted and rejected components, and Grover reflections recombine
them~\cite{grover1996search,brassard2002amplitude,boyer1998tight}.  After a
finite coherent evolution, the final success probability peaks when the
success probability before interference reaches an analytically determined
nonunit value.  Training
toward this interior target, a preferred probability strictly between zero
and one, can improve the final result while retaining
multiple reasoning paths, thereby mitigating policy collapse in this setting.

We train and evaluate both models on a $2\times3$ sliding puzzle
[Fig.~\ref{fig:puzzle}], a slow-thinking benchmark introduced in
Ref.~\cite{hu2025freezing}, in which solving a question
requires a sequence of interdependent moves.  The target configuration is
supplied, but the model must discover a valid trajectory that reaches it, so an
exact verifier can evaluate every candidate trajectory.  Generic question
answering would instead require a learned verifier or value model to mark
candidate outcomes coherently.  The trained models retain
branching ensembles of reasoning paths [Fig.~\ref{fig:paradigm}(b),(c)].  For
a matched comparison of inference computation, the quantum and classical
methods receive the same number of applications of the reasoning circuit:
the classical method uses them to sample independent trajectories,
whereas the quantum method uses them for state preparation and Grover
interference.  At four Grover rounds, QuCoNet and the neural quantum AI model
achieve respectively $3.2$ and $3.9$ times the held-out accuracy of the
strongest classical controls, where held-out accuracy is the mean success
probability on question--answer pairs that were not used for training.  Over
the tested range, the number of question--answer pairs supported by a policy
of fixed size, trained separately at each budget, also grows faster with the
amplification budget (the number of
reasoning-circuit applications) than with matched classical repetition.  These results establish a trainable, Grover-based realization of
path-integral slow thinking, in which the interior target defined by Grover
inference preserves multiple reasoning paths while the amplification improves
verified performance.

In the present framework, quantum dynamics is the trainable reasoning process
itself.  Running more Grover rounds increases the inference computation per
question, and a policy of fixed size trained for a larger round budget
supports more question--answer pairs.  The training objective is the
same quantity computed during inference: the success probability after a
chosen number of rounds.  Grover interference is therefore not appended after
training to a policy optimized for direct measurement: the policy is trained
to prepare the trajectory distribution on which its later coherent reasoning
will act.

The remainder of this Article is organized as follows.
\hyperref[sec:trajectories]{Section~\ref*{sec:trajectories}} develops the
statistical-trajectory and path-integral descriptions of classical and quantum
reasoning.  \hyperref[sec:models]{Section~\ref*{sec:models}} introduces the
sliding-puzzle benchmark and the table-based and neural quantum AI models.
\hyperref[sec:classical-rlvr]{Sections~\ref*{sec:classical-rlvr}} and
\hyperref[sec:quantum-rlvr]{\ref*{sec:quantum-rlvr}} explain RLVR and policy
collapse, then construct quantum RLVR through Grover interference and derive
its finite-depth training target.
\hyperref[sec:results]{Section~\ref*{sec:results}} compares the training
objectives, held-out performance, path diversity, and scaling with inference
depth.  \hyperref[sec:conclusion]{Section~\ref*{sec:conclusion}} concludes with
extensions to learned verifiers and the requirements for physical realization.

\section{Classical and Quantum Reasoning Trajectories}
\label{sec:trajectories}

Let $\theta$ denote the trainable model parameters.  For an autoregressive LLM,
a reasoning trajectory $\gamma=(z_1,\ldots,z_T)$ has probability
\begin{equation}
P_{\theta}[\gamma|x]
=\prod_{t=1}^{T}\pi_{\theta}(z_t|x,z_{<t})
=e^{-I_{\mathrm{LLM}}[\gamma|x]} .
\label{eq:llm-path-measure}
\end{equation}
Here $\pi_\theta(z_t|x,z_{<t})$ is the normalized conditional probability of
the next reasoning state, and $z_{<t}=(z_1,\ldots,z_{t-1})$ is the complete
prefix.  Thus $I_{\mathrm{LLM}}[\gamma|x]=-\log P_{\theta}[\gamma|x]$ is a
dimensionless information action, or trajectory surprisal.
Equation~\eqref{eq:llm-path-measure} makes no Markov approximation: each
conditional may depend on the complete prefix.  A quantum reasoning circuit
associates a complex transition amplitude $\alpha_{\theta}(z_t|x,z_{<t})$
with each conditional, so the path amplitude is
$A_{\mathrm Q}[\gamma|x]=\prod_t\alpha_{\theta}(z_t|x,z_{<t})$, whose phase is
written schematically as $S_{\mathrm Q}[\gamma|x]/\hbar$.  Let
$|\gamma\rangle$ denote the basis state that records the complete trajectory,
and let $M_y$ be the element of a positive operator-valued measure (POVM)
associated with answer $y$.  The answer probability is then
\begin{equation}
P_{\mathrm Q}(y|x)=
\sum_{\gamma,\gamma'}A_{\mathrm Q}[\gamma|x]
A_{\mathrm Q}^{*}[\gamma'|x]
\langle\gamma'|M_y|\gamma\rangle .
\label{eq:quantum-history-readout}
\end{equation}
The $\gamma=\gamma'$ terms form a classical mixture of histories; nonzero
off-diagonal terms permit interference whenever the measurement coherently
recombines the corresponding histories.  In the Grover realization studied
here, fresh which-path records keep individual trajectories orthogonal.  No
pairwise interference between these records contributes to the success
probability or to the path statistics reported below.  Instead, the verifier
reflection and the reflection about the prepared state act coherently on the
collective accepted and rejected components, as described below.

\section{Sliding-Puzzle Benchmark and Quantum Reasoning Models}
\label{sec:models}

\begin{figure*}[t]
  \includegraphics[width=\textwidth]{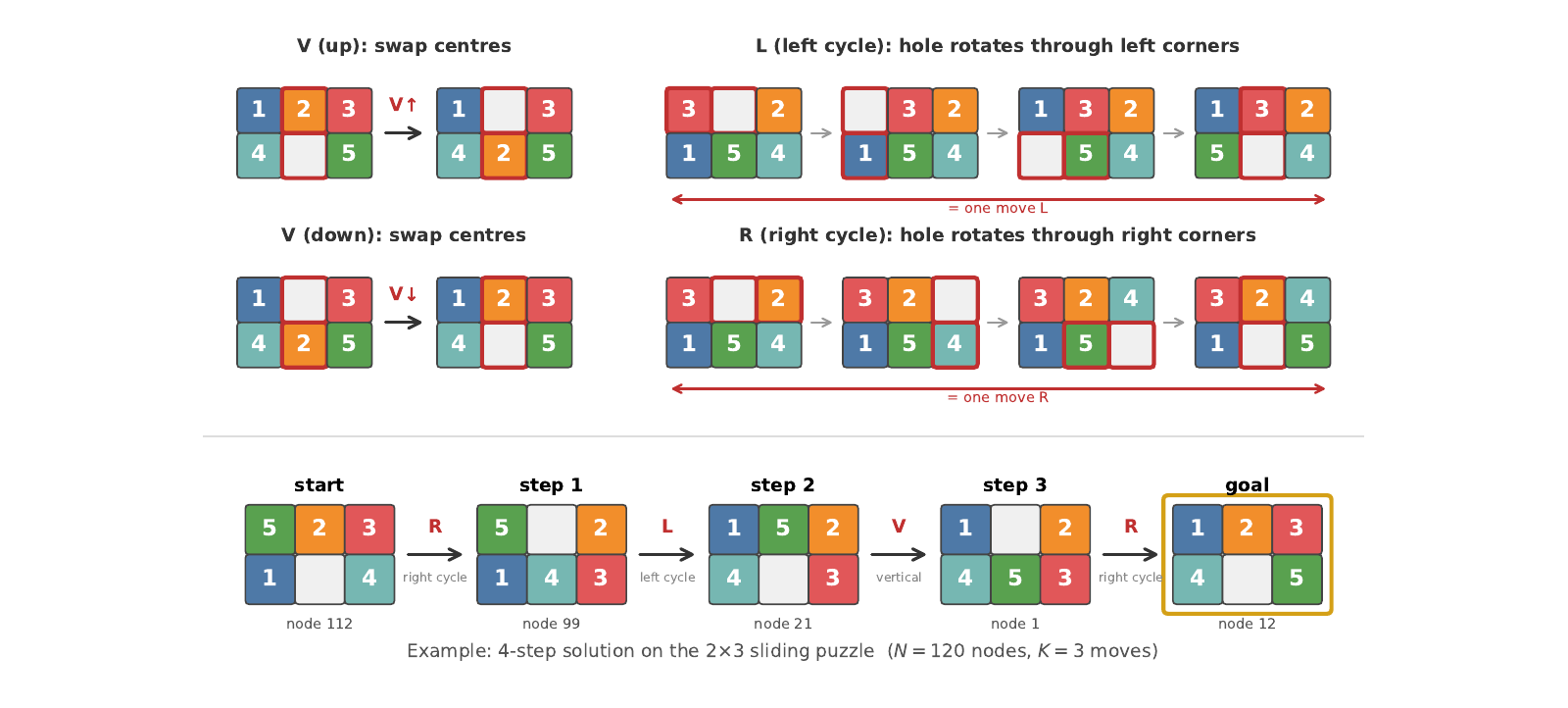}
  \caption{\textbf{The sliding puzzle as a network benchmark for slow thinking.}
  Top: the three reversible graph moves.  $V$ swaps the two central cells;
  $L$ and $R$ move the hole through the left or right corners using three
  elementary tile slides.  Each complete $L$ or $R$ cycle counts
  as one action and one edge of the configuration graph.  Restricting the hole
  to the middle column and the tile permutation to one parity class gives
  $N=120$ reachable configurations, each with $K=3$ available actions.
  Bottom: a question--answer pair specifies start and target nodes in this
  graph.  The target is supplied to the model, whose output is a trajectory
  connecting the two nodes.  The
  example requires four actions.  The verifier accepts on first arrival at the
  target within the reasoning horizon $M=8$; the node numbers label graph
  configurations.}
  \label{fig:puzzle}
\end{figure*}

We make the trajectory framework concrete by turning a goal-directed
$2\times3$ sliding puzzle~\cite{hu2025freezing} into a finite configuration
network [Fig.~\ref{fig:puzzle}].  Each allowed arrangement of the five numbered tiles
and one hole is a node, and each reversible move is an edge.  We restrict the
hole to the middle column and define three moves: $V$ exchanges the two central
cells, whereas $L$ and $R$ carry the hole around the left or right corners
through three elementary tile slides.  Each complete $L$ or $R$ cycle counts
as one action and hence one network edge.  Restricting the tile permutation to
one parity class produces a $3$-regular graph with $N=120$ nodes and $K=3$
available actions at every node (construction in the Supplemental Material).

A question--answer pair $x_i=(Q_i,A_i)$ is a pair of nodes in this network:
$Q_i$ is the start configuration and $A_i$ is the target configuration, chosen
at graph distance $3$--$6$ from $Q_i$.  The target $A_i$ is supplied to the
model; the nontrivial output is an ordered trajectory
$\gamma_i=(a_1,\ldots,a_T)$ with a prefix that carries the configuration from
$Q_i$ to $A_i$.  We allow
at most $M=8$ actions, and an exact binary verifier accepts a trajectory on
its first arrival at $A_i$ within this horizon.  The four-action
example in Fig.~\ref{fig:puzzle} makes the slow-thinking character explicit:
the answer can be obtained only through a sequence of interdependent actions,
not from any single intermediate state.  For each question pool, the first
$B$ question--answer pairs form the training set and $64$ pairs disjoint from
the training set form the held-out set; model snapshots are selected using the training set only.
Throughout, \emph{accuracy} denotes the mean final success probability over
the specified set of question--answer pairs.

This network abstraction is not specific to sliding puzzles.  Many multistep
tasks can be represented as navigation through an implicit state network:
nodes may denote partial mathematical derivations, partial proofs, intermediate
programs, or physical configurations, while edges denote valid algebraic
transformations, inference rules, code edits, or control actions.  Such
networks are generally enormous, history dependent, and not explicitly
available to the model.  They nevertheless share the ingredients isolated
here: a starting condition, branching sequences of actions, delayed
verification, and potentially many trajectories to the same outcome.  The
sliding puzzle is therefore a controlled benchmark of AI reasoning: it retains
these structural features while providing exact transitions and an exact
verifier, so trajectory exploration, policy collapse, and interference can be
measured without ambiguity.

\begin{figure}[!ht]
  \centering
  \includegraphics[width=0.88\columnwidth]{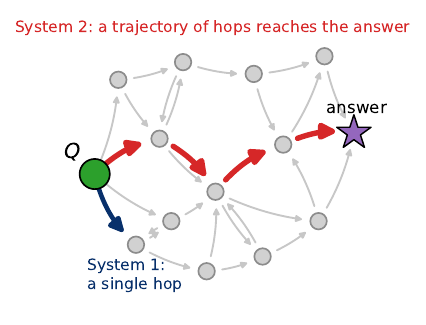}
  \caption{\textbf{Slow thinking as a walk on a concept network.}
  Nodes denote coarse-grained concepts, later given an operational definition
  as effective predictive states, and directed edges denote learnable local
  transitions.  A single hop is a fast System-1
operation~\cite{kahneman2011thinking}.  Repeatedly
  applying the same local dynamics constructs a trajectory from a question
  $Q$ through intermediate states to an answer, realizing System-2 slow
  thinking as evolution in time.  Red arrows highlight one such trajectory.}
  \label{fig:conet}
\end{figure}

This state-network view defines the concept-network model (CoNet)
[Fig.~\ref{fig:conet}].  CoNet was introduced in the Learning-at-Criticality
framework as a minimal dynamical model for studying learning transitions and
generalization in LLMs~\cite{cai2025criticality}.  Its original construction
coarse-grained the nearly deterministic, low-entropy token sequences into
concept nodes, represented the high-entropy choices between them by learnable links,
and assumed Markovian dynamics among the resulting nodes.  A reasoning episode
was then a walk from a question node to an answer node, with reinforcement
learning updating the local transition probabilities.

The subsequent work of Ref.~\cite{hu2025freezing} applied and extended this
framework to explain how RLVR can produce long chains of thought and emergent
slow thinking.  That analysis associated the onset of slow thinking with an
inverse-tree-freezing transition, in which reinforcement stabilizes extended
multistep trajectories.  It also placed the concept-state construction on an
operational footing through predictive state variables (PSVs).  Two token
histories belong to the same PSV when they induce indistinguishable
conditional distributions over future continuations.  The resulting
equivalence classes are effective states on which the dynamics is Markovian,
replacing the earlier coarse-graining assumption with an operational
definition.  CoNet is therefore not an LLM architecture, but an effective
dynamical description of reasoning unfolded in time.

Here we use the finite configuration network above as a concrete CoNet and
promote its stochastic routing dynamics to coherent quantum dynamics.  This
extension provides a minimal quantum AI model for studying quantum RLVR.  Its
table-based realization, QuCoNet, replaces the local probability vector by a
trainable $3\times3$ unitary, called the coin operator.  Each step of the
resulting coined quantum walk alternates this local redistribution of
amplitude with a reversible shift to the selected neighboring state, thereby
propagating amplitudes coherently over action sequences~\cite{dernbach2019qwnn}.
The neural quantum AI model instead uses a
parameter-sharing network to generate the coin from the question and the
preceding action sequence.  This removes the need for a separate parameter
table at every node and allows the local quantum dynamics to depend on the
complete reasoning prefix.  The classical CoNet and both quantum models route
weight through intermediate reasoning states; they differ in how the
trajectory distribution is represented, evolved, and measured.

In either quantum model, each move is recorded in a fresh auxiliary register,
called a which-path register
[Fig.~\ref{fig:paradigm}(a)].  The trainable reasoning circuit
$\mathcal A_\theta$ then prepares a coherent superposition
$|\psi_i\rangle=\mathcal A_\theta|0\rangle$ over all
$K^M=3^8$ action sequences.  Here $|0\rangle$ denotes the initialized circuit
registers, and the dependence of $\mathcal A_\theta$ on question $x_i$ is
implicit.  Let $P_{\mathrm{acc},i}$ project onto the trajectories accepted by
the verifier for question $i$.  Their total weight before interference is
\begin{equation}
 p_i(\theta)
 =\langle\psi_i|P_{\mathrm{acc},i}|\psi_i\rangle
 =\sum_{\gamma\in\Gamma_i^{\mathrm{acc}}}P_\theta[\gamma|x_i],
 \label{eq:base-success}
\end{equation}
where $\Gamma_i^{\mathrm{acc}}$ is the set of accepted trajectories and
$P_\theta[\gamma|x_i]\equiv|A_{\mathrm Q}[\gamma|x_i]|^2$ is the path weight
obtained by measuring the which-path records before amplification.  Measuring these
records samples the corresponding classical distribution; retaining them
coherently allows the accepted and rejected components to be recombined before
measurement.  In the present study, we evaluate Eq.~\eqref{eq:base-success} and its
gradients by exact classical enumeration, thereby isolating coherent
inference as the quantum resource rather than conflating it with faster
training.  Full-state propagation in the
$N K^M=120\times3^8=787{,}320$-dimensional Hilbert space of positions and path
records
independently verifies the coherent circuit construction (Supplemental
Material).

\section{RLVR and Policy Collapse}
\label{sec:classical-rlvr}

Equation~\eqref{eq:base-success} provides the interface between the reasoning
model and its training objective.  For each training pair $i$, the parameters
$\theta$ determine the weights of all reasoning trajectories, the verifier
identifies the accepted trajectories, and their total weight gives
$p_i(\theta)$.  Training then differentiates a loss constructed from these
success probabilities and updates the same parameters:
\[
 \theta\ \longrightarrow\ P_\theta[\gamma|x_i]
 \ \longrightarrow\ p_i(\theta)
 \ \longrightarrow\ \mathcal L(\theta)
 \ \longrightarrow\ \theta .
\]
This procedure applies to all three trainable policies: the transition weights
of classical CoNet, the local coin tables of QuCoNet, and the shared parameters
of the neural quantum AI model.  In our calculations, every
$p_i(\theta)$ and its gradient are evaluated exactly, so the comparison below
differs in the training objective and the subsequent inference dynamics, not
in the accuracy of a sampling-based gradient estimate.

One widely used practical realization of RLVR is group relative policy
optimization (GRPO)~\cite{shao2024deepseekmath}.  GRPO samples several
complete trajectories for the same question and compares each terminal reward
with the mean reward of the group.  Trajectories scoring above the group mean
are made more probable, while those scoring below it are made less probable.
The group therefore supplies the reference value without a separately trained
value model, or critic.

In practical GRPO, this learning signal is estimated from sampled groups of
trajectories.  Because the present benchmark has a finite trajectory space, we
can instead compute the complete sum over trajectories and its gradient
exactly.  Assigning unit reward to an accepted trajectory and zero to a
rejected one makes this sum, $p_i(\theta)$, the expected reward of a single
inference attempt, so training minimizes
\begin{equation}
 \mathcal L_{\mathrm{RLVR}}(\theta)=-\sum_i p_i(\theta).
 \label{eq:classical-rlvr}
\end{equation}
The loss decreases whenever any $p_i$ increases and therefore pushes every
training question toward unit success.  It depends only on the total accepted
weight and is indifferent to how that weight is divided among distinct
successful trajectories.  Consequently, it supplies no counterpressure when
one successful route grows at the expense of alternatives.

This indifference becomes a tendency toward \emph{policy collapse} when a
shared policy with finite capacity must serve many questions.  Increasing the
probability of an already successful route can improve the reward more easily
than maintaining several weakly sampled alternatives.  In practical RLVR,
sampling adds further positive feedback: routes that succeed early are
observed and reinforced more often, whereas competing routes receive fewer
updates.  Training can therefore obtain high single-attempt accuracy while
losing the diversity needed to explore unfamiliar problems.  The loss does
not mathematically force the policy to place all probability on a single
trajectory; rather, it contains no mechanism that protects alternative paths.  The collapse observed
under exact evaluation in Fig.~\ref{fig:collapse}(b) also shows that the effect
in our benchmark is not solely a finite-sampling artifact.

Classical test-time repetition improves the probability of finding an answer
but does not change this training pressure.  If each independent
attempt succeeds with probability $p$, at least one of $k$ attempts succeeds with
probability
\begin{equation}
 C_k(p)=1-(1-p)^k,
 \qquad \frac{dC_k}{dp}=k(1-p)^{k-1}>0 .
 \label{eq:classical-repetition}
\end{equation}
This success probability is still monotonic in $p$.  Training for repeated
sampling therefore continues to favor the largest possible single-attempt
success probability and
creates no interior point at which the policy should stop concentrating its
probability.  The same statement applies to adaptive procedures that observe
only the final binary verdict of each completed trajectory (Supplemental
Material).  It does not apply to methods that inspect intermediate states,
modify the policy between attempts, or explicitly reward diversity.  Within
this outcome-only setting, changing a classical policy into a quantum
parameterization is insufficient: if the prepared quantum state is measured
directly, the measurement samples the same trajectory probabilities and the
objective is the same monotonic one.  A different training pressure therefore requires a
coherent operation that changes the success probability before measurement.

\section{Quantum RLVR through Grover Interference}
\label{sec:quantum-rlvr}

Quantum reinforcement learning has used quantum dynamics in several distinct
roles.  Quantum projective simulation replaces a stochastic walk through the
agent's episodic memory by a quantum walk, making coherence part of the
agent's internal deliberation before it selects an
action~\cite{paparo2014agents,katabarwa2017hamiltonian,
franceschetto2024projective}.  Parametrized quantum circuits instead learn a
map from an observed state to an action
distribution~\cite{jerbi2021parametrized}, following earlier quantum
reinforcement-learning frameworks~\cite{dong2008qrl}.  Other algorithms use
amplitude estimation or Grover search to accelerate policy evaluation, policy
improvement, or inference~\cite{wiedemann2023qpi,zeng2019generative}.  Recent
quantum Markov decision processes go further by coherently generating
multistep state--action trajectories, computing their returns, and amplifying
the high-return sector~\cite{su2025qmdp,su2025grover}.  These works are direct
precursors: quantum deliberation, trainable quantum policies, coherent
trajectory propagation, and amplitude-amplified selection all existed before
the present study.

Our distinction is therefore not the mere presence of a quantum policy or a
multistep coherent process.  We ask whether a learned model of fixed size can convert
additional coherent inference time into increased reasoning capacity, in
direct analogy with test-time scaling in LLMs.  The policy first prepares a
superposition of complete reasoning trajectories, which undergoes
verifier-controlled Grover evolution before any outcome is measured.  The
number of Grover rounds is then an inference-time resource: it changes the
computation performed at inference rather than the number of trainable
parameters.  Crucially, the same finite-depth dynamics also defines the
training objective.  The model is not optimized for the directly measured
base probability $p_i(\theta)$ and then accelerated by a quantum subroutine; it
is optimized for the composite success map $G_{\le q}[p_i(\theta)]$ that will
determine its success at inference.  Coherence thus changes both how the model reasons
and what trajectory distribution it is trained to represent.  This joint
dependence of inference and training is what we call quantum slow thinking.

The essential step is to postpone measurement until the verifier has acted
coherently.  Operationally, one inference run begins when the reasoning
circuit $\mathcal A_\theta$ prepares a superposition of complete reasoning
trajectories.  Each Grover round then performs two operations.  First, a
coherent verifier applies a minus sign to trajectories whose final outcomes
are accepted, without revealing which outcome occurred.  Second, the
sequence $\mathcal A_\theta^\dagger$, then a phase flip about the initial state
$|0\rangle$, then $\mathcal A_\theta$ reflects the state about the prepared
superposition.  The verifier alone changes no probability; the second
reflection converts its phase contrast into amplitude transferred from the
rejected sector to the accepted sector.  Measurement is performed only after
the chosen number of complete rounds.

This sequence is Grover amplitude amplification applied to
reasoning~\cite{grover1996search,brassard2002amplitude}.  For
one question, we suppress the index $i$ and decompose the state prepared by
$\mathcal A_\theta$ into collective components in the two orthogonal
subspaces, or sectors, accepted and rejected by the verifier,
\begin{equation}
 \begin{aligned}
 |\psi\rangle
 &=\sqrt p\,|\psi_{\mathrm{acc}}\rangle
 +\sqrt{1-p}\,|\psi_{\mathrm{rej}}\rangle,\\
 p&=\sin^2\vartheta,\qquad 0\le\vartheta\le\frac{\pi}{2},
 \end{aligned}
 \label{eq:good-bad-main}
\end{equation}
where $|\psi_{\mathrm{acc}}\rangle$ and
$|\psi_{\mathrm{rej}}\rangle$ are the normalized collective components, and
$p$ is the success probability before interference.  The verifier reflection
$R_{\mathrm{acc}}=I-2P_{\mathrm{acc}}$ gives the two components opposite phases but
does not by itself change their probabilities.  Reflection about the prepared
state,
$R_\psi=2|\psi\rangle\!\langle\psi|-I
=\mathcal A_\theta(2|0\rangle\!\langle0|-I)\mathcal A_\theta^\dagger$,
converts this phase difference into a redistribution of accepted and rejected
weight.  Together the two reflections form one Grover round.  Which-path
records keep individual trajectories orthogonal, while the complete Grover
round coherently rotates the collective accepted and rejected components.  No
pairwise interference between individual path records is required.  This is a
two-sector coarse graining of the path integral in
Eq.~\eqref{eq:quantum-history-readout}: all accepted histories form one sector,
all rejected histories form the other, and their relative phase becomes
observable when the two sectors are coherently recombined.

Geometrically, the evolution remains in the two-dimensional plane spanned by
$|\psi_{\mathrm{acc}}\rangle$ and $|\psi_{\mathrm{rej}}\rangle$.  If the initial accepted amplitude is
$\sin\vartheta$, each complete Grover round rotates the state by
$2\vartheta$ toward the accepted sector.  After $r$ rounds, the accepted
amplitude is therefore $\sin[(2r+1)\vartheta]$.

These $r$ rounds involve $q=2r+1$ applications of the
reasoning circuit $\mathcal A_\theta$ or its inverse
$\mathcal A_\theta^\dagger$: one application prepares the state, and each
round adds one application in each direction.  The final success probability is
\begin{equation}
 G_q(p)=\sin^2\!\left(q\arcsin\sqrt p\right),
 \qquad q=2r+1 .
 \label{eq:grover-map}
\end{equation}
Unlike $C_k(p)$, this map is not monotonic.  Its first maximum reached from
$p=0$ occurs at the nonunit base probability
\begin{equation}
 \pstar(q)=\sin^2\!\left(\frac{\pi}{2q}\right),
 \qquad \pstar(3)=\frac14 .
 \label{eq:pstar}
\end{equation}
Below this value, increasing $p$ improves the final result; immediately above
it, the Grover rotation overshoots and the gradient reverses.  The inference
dynamics therefore supplies an interior training target rather than pushing
$p$ monotonically toward one.

Training proceeds exactly as in the classical loop above, except that the
verifiable reward is evaluated after coherent inference rather than directly
after state preparation.  This is the sense in which the procedure is quantum
RLVR: the correctness criterion is still a terminal verifier, but quantum
dynamics transforms the success probability on which the policy is trained.
When at most $q$
applications of the reasoning circuit are available, we stop at the Grover
round nearest the first maximum and denote the resulting probability by
$G_{\le q}(p)$; the exact rule is given in the Supplemental Material.  We then
minimize
\begin{equation}
 \mathcal L_{\mathrm Q}(\theta)
 =-\sum_iG_{\le q}\!\left[p_i(\theta)\right].
 \label{eq:quantum-rlvr}
\end{equation}
The gradient passes through $p_i(\theta)$ to the same table or neural
parameters as before.  Unlike Eq.~\eqref{eq:classical-rlvr}, however,
Eq.~\eqref{eq:quantum-rlvr} trains each question toward $\pstar(q)$.  The
reasoning policy need only place this finite weight on accepted trajectories;
the subsequent coherent evolution amplifies it to unit success.  Thus the
inference dynamics determines both what the model learns and how the learned
trajectory weights are converted into an answer.

A longer coherent evolution can amplify a smaller initial accepted weight:
for small $p$, $G_q(p)=q^2p+O(p^2)$.  Reaching constant success consequently
requires $O(1/\sqrt p)$ coherent applications of the reasoning circuit,
whereas independent verified sampling requires $O(1/p)$ attempts.  Here and
below, these are query counts in the small-$p$ limit, rather than total
hardware costs.  This
quadratic relation explains why a longer coherent evolution can support a
smaller initial success probability.  This scaling also marks the scope of the
present mechanism: the interior target follows specifically from Grover
interference, coherent access to $\mathcal A_\theta$ and
$\mathcal A_\theta^\dagger$, and a verifier that
can mark accepted trajectories without measurement.  Other coherent inference
maps may generate different training objectives.  Exact classical evaluation
in the present experiments isolates this coupling between training and
inference; implementing the same loop on quantum hardware additionally
requires coherent evaluation of the verifier and a suitable method for
estimating gradients.

\section{Results}
\label{sec:results}

\subsection{Grover-based training preserves path diversity}

\begin{figure*}[t]
  \includegraphics[width=\textwidth]{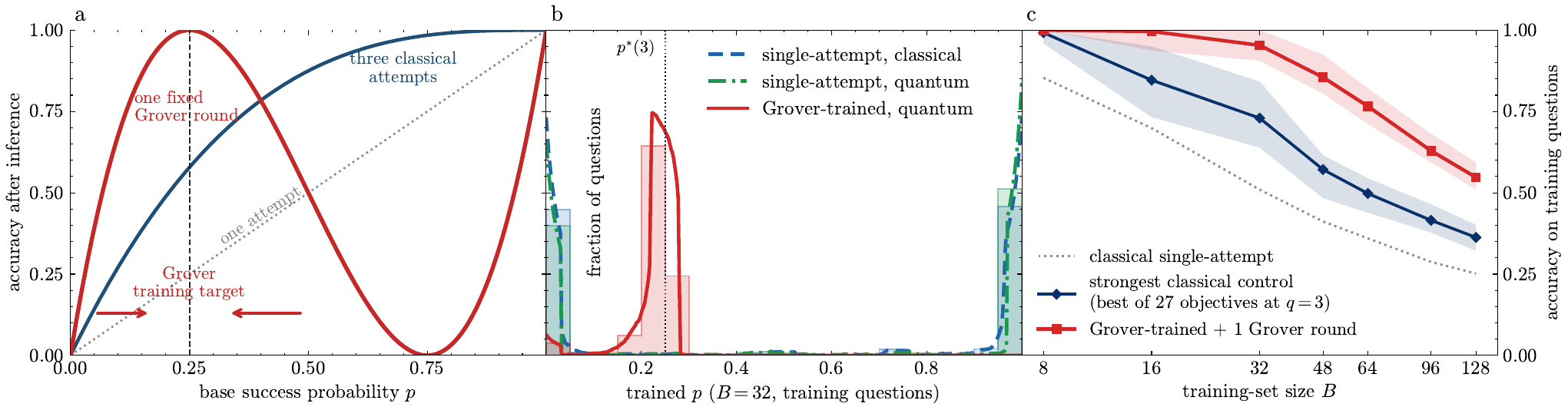}
  \caption{\textbf{Grover-based training produces an interior success target
  and preserves accuracy at matched inference cost.}
  (a)~The classical single-attempt objective increases monotonically with the
  pre-amplification success probability $p$.  The Grover success probability
  instead reaches its first maximum at $p=\pstar(q)$; arrows show the
  corresponding training directions.  (b)~Distributions of $p_i$ for $B=32$,
  computed exactly over $32$ question-pool seeds.  Classical single-attempt training drives
  both classical and quantum models toward $p_i=0$ or $1$, whereas
  Grover-based training concentrates them near $\pstar(q)$.  Curves are
  smoothed histograms of the displayed data.  (c)~Mean training accuracy
  versus the number $B$ of training question--answer pairs at $q=3$.
  Grover-based training exceeds the best result among $27$ classical controls
  granted the same three circuit applications, at every tested $B$.  Bands span the
  minimum and maximum values across question-pool seeds; control definitions and plotting details are
  given in the Supplemental Material.}
  \label{fig:collapse}
\end{figure*}

We first tested whether training produced the interior success probability
predicted by Eq.~\eqref{eq:pstar}.  Under the classical single-attempt
objective, the distribution of the pre-amplification success probability
$p_i$ became bimodal.  Questions that were learned accumulated near $p_i=1$,
whereas the remaining questions accumulated near $p_i=0$
[Fig.~\ref{fig:collapse}(b)].  The same behavior occurred in CoNet and in both
quantum models, showing that quantum parameterization alone did not prevent
policy collapse.  Training with the Grover objective instead produced a peak
near the predicted value $\pstar(q)$ for every tested inference depth $q$.
Thus the analytic optimum of the Grover dynamics was also a stable target of
the training dynamics.

To diagnose the training objective, we used the adaptive rule $G_{\le q}$,
which allowed at most $q$ circuit applications and stopped at the Grover
round nearest the first maximum for each question.  It yielded mean training accuracy above
$0.95$ at every tested $q$.  All held-out comparisons below instead used a
fixed-depth protocol that applied the full $q$ circuit operations without
question-specific stopping.  Applying this fixed depth to the training set
reduced its accuracy to $0.77$ at $q=9$, because questions with a larger
initial $p_i$ reached their maximum earlier and were then rotated past it
(Supplemental Material).

The two objectives also produced different ensembles of successful paths.
Classical single-attempt training concentrated most of the successful
probability on one route [Fig.~\ref{fig:paradigm}(b)].  Grover-based training
retained several routes [Fig.~\ref{fig:paradigm}(c)].  The accuracy-weighted
inverse participation ratio (IPR), which counts the effective number of
successful paths, increased from $1.6$ at $q=3$ to $5.7$ at $q=9$ for $B=32$
(Supplemental Material).  This quantity measures path diversity, not pairwise
interference between the recorded paths.  The control analysis below traces
this retention to the interior target itself: a classical loss trained toward
the same target preserves at least as much diversity.

\subsection{Performance at matched inference cost}

We compared quantum and classical inference using the same number $q$ of
applications of the reasoning circuit $\mathcal A$.  A quantum inference with
$r$ Grover rounds used $q=2r+1$ applications: one for state preparation and
two per round.  The matched classical method used the same $q$ applications
for $q$ independent trajectories.  Thus $q$ counts forward or inverse uses of
the reasoning circuit; verifier evaluations and reflections are reported
separately in the Supplemental Material.  This comparison isolates the
inference mechanism; it does not include the hardware costs of reversible
computation, trajectory memory, or fault tolerance.

The performance difference increased with inference depth
[Table~\ref{tab:ladder}].  For $B=32$ and $q=9$, corresponding to four Grover
rounds, table-based QuCoNet reached held-out accuracy $0.407$, and the neural
quantum AI model reached $0.496$.  The strongest classical control reached $0.126$
at the same inference cost.  The two quantum AI models therefore achieved
$3.2$ and $3.9$ times the held-out accuracy of the classical control.  At
$q=3$, QuCoNet remained below the classical control; from $q=5$ onward both
quantum models were higher.

Each entry in Table~\ref{tab:ladder} used the same fixed value of $q$ for every
question and did not use $p_i$ to choose when to stop.  The increase was a
property of training, rather than only of applying a deeper circuit: when
models trained at different values of $q$ were evaluated at a common depth,
the held-out accuracy still increased with the value of $q$ used during
training (Supplemental Material).

\begin{table}[t]
\caption{\textbf{Held-out accuracy at matched inference cost.}
The learned systems were trained on $B=32$ question--answer pairs and evaluated with
$q$ circuit applications.  Entries are means over independent replicas;
parentheses give replica-to-replica standard deviations in the last digits.
The two quantum-model rows use $32$ replicas.  For the tabular models, replicas
vary the question pools while fixing the optimization seed; for the neural
model, the question-pool and model-initialization seeds vary together.  The
classical row reports the best result among $27$ classical controls at each
$q$ and data split, using $32$ replicas
per control.  Each quantum model used the same fixed $q$ for every question.
The uniform-policy row provides a diversity-preserving reference rather than a
learned solution.  Complete control definitions and training-set results are
given in the Supplemental Material.}
\label{tab:ladder}
{\footnotesize\setlength{\tabcolsep}{2pt}
\begin{ruledtabular}
\begin{tabular}{@{}lcccc@{}}
applications $q$          & 3 & 5 & 7 & 9\\
\midrule
QuCoNet               & 0.042(20) & 0.147(34) & 0.272(39) & 0.407(45)\\
neural quantum AI     & 0.073(17) & 0.193(25) & 0.340(30) & 0.496(43)\\
best classical        & 0.071(23) & 0.091(27) & 0.109(24) & 0.126(25)\\
uniform policy + Grover
                        & 0.077(4)  & 0.204(9)  & 0.367(15) & 0.544(20)
\end{tabular}
\end{ruledtabular}
}
\end{table}

\begin{figure*}[t]
  \includegraphics[width=\textwidth]{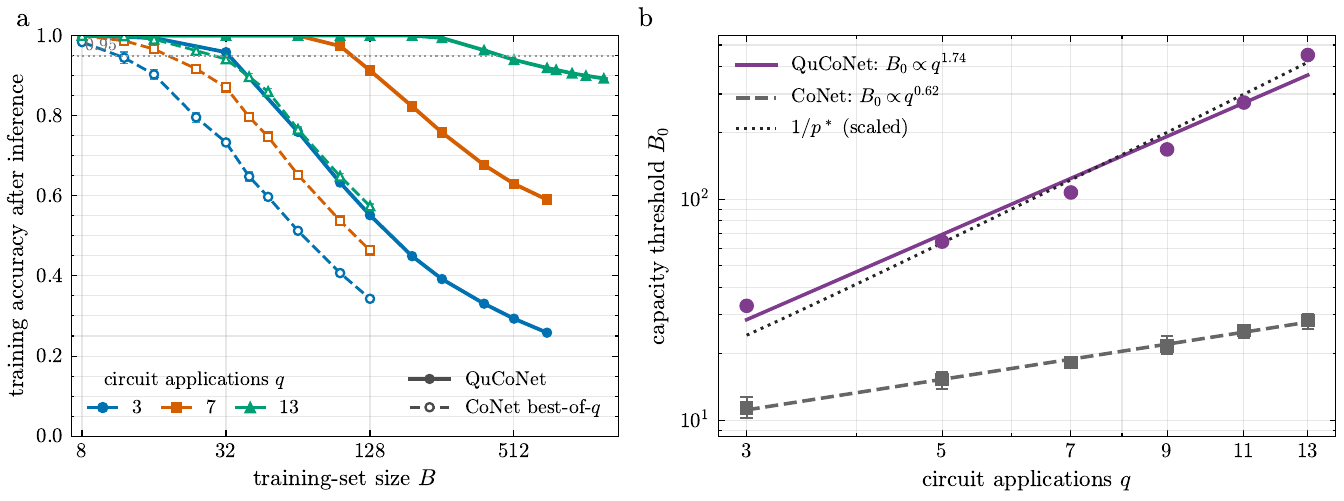}
  \caption{\textbf{Coherent inference increases the number of
  question--answer pairs supported by a policy of fixed size trained at each
  budget.}
  (a)~Mean training accuracy versus the number $B$ of training
  question--answer pairs at
  matched values of $q$.  Solid curves show Grover-based training and
  inference; dashed curves show classical best-of-$q$ (the matched protocol
  using $q$ independent trajectories) training and inference
  with the same $720$-parameter table-based policy.  The dotted line at $0.95$
  defines $B_0(q)$ as the largest supported training-set size.  Error bars are
  central $68\%$ bootstrap intervals from $16$ shared question pools.  Panel
  (a) shows $q=3,7,13$; all six odd values from $q=3$ to $13$ are used in
  panel (b).  (b)~The extracted $B_0(q)$ values and finite-range power-law
  fits.  The coherent result approximately follows $1/\pstar(q)$, whereas the
  classical capacity increases more slowly.  Complete sampling and fitting
  procedures are given in the Supplemental Material.}
  \label{fig:results}
\end{figure*}

We next asked whether a longer coherent evolution allows a policy of the
same size to
support more question--answer pairs.  Figure~\ref{fig:results} summarizes this
capacity comparison; its quantitative interpretation follows after the
control analysis below.

\subsection{Controls isolate joint training and interference}

At each value of $q$, the classical reference in Table~\ref{tab:ladder} was the
best result among $27$ training and inference controls.  These included
single-attempt and repeated-sampling objectives, entropy regularization, and
controls using the same table-based policy as QuCoNet (Supplemental Material).
All of these controls observe only the final verified outcome of complete
trajectories, whether estimated from sampled verdicts or computed as its
exact expectation (Supplemental Material); classical algorithms that exploit the known transition graph,
such as breadth-first search, solve this benchmark exactly and lie outside
the access model compared here.
At $B=32$ and $q=3$, Grover-trained QuCoNet reached training accuracy $0.953$,
compared with $0.729$ for the strongest classical control
[Fig.~\ref{fig:collapse}(c)].  A control applying Grover amplification only
after classical training reached $0.919$.  Its trained probabilities spread
around the interior value instead of concentrating on it, so the single round
converted them imperfectly.

We also asked whether a classical loss could train the policy directly toward
the same numerical target.  A capped loss stopped rewarding a question once
$p_i$ exceeded $\pstar(3)=0.25$, but did not penalize probabilities that moved
above the target.  After one fixed-depth Grover round, this control reached
training accuracy $0.706$ (Supplemental Material).  A symmetric loss that penalized deviations on both sides of
$\pstar(q)$ reproduced the held-out accuracies of Grover-based training when
it was paired with the same coherent inference.  It also preserved at least
as much path diversity, with accuracy-weighted IPR $9.3$ against $5.8$ at
$q=9$ on shared question pools (Supplemental Material), so the retained
diversity follows from the interior target rather than from the Grover form
of the loss.  Thus classical optimization
could reproduce the numerical target, but the final performance still
required the quantum amplification associated with that target.  For example,
three independent samples at $p=0.25$ succeed with probability $0.578$,
whereas one Grover round reaches unit success.  Further controls are reported
in the Supplemental Material.

The untrained uniform table policy provided a reference for the path diversity removed by
training [Table~\ref{tab:ladder}, last row].  Followed by
Grover amplification, it reached the highest held-out accuracy because it had not
specialized to the training questions.  However, it performed nearly
identically on the training and held-out sets and therefore represented no
learned solution.  Its success probabilities also lie so deep in the
small-$p$ tail that no question reaches its Grover maximum within the tested
budgets, so this reference cannot bring any designated question set to the
near-unit accuracy that trained policies reach at $q=3$ (Supplemental
Material).  At $q=9$, the trained quantum models retained $75$--$91\%$
of this reference accuracy on held-out questions.  The strongest matched
classical control retained just under a quarter.

These comparisons were insensitive to the stopping convention and persisted
across the full range of $B$ and on a second random-regular graph family.  At
$B=32$, allowing the stopping rule to use $p_i$ changed the held-out values in
Table~\ref{tab:ladder} by less than $0.012$.  In a size scan, the untrained
model's lead over Grover-trained QuCoNet decreased from $1.7\times$ at $N=120$
to within question-pool variation at $N=480$, while the trained model's lead
over the classical control increased.  On the harder distance-$7$, $M=9$
questions, the trained neural quantum AI model matched the untrained reference
(Supplemental Material).

\subsection{Reasoning capacity}

The smaller target $\pstar(q)$ at larger $q$ reduced the
pre-amplification success probability required for each training question.
We therefore tested whether policies of this fixed size, trained separately
at each budget $q$, could support more question--answer
pairs as that budget increased.  For each $q$, we defined
$B_0(q)$ as the largest training-set size $B$ for which the mean accuracy after
adaptive Grover inference ($G_{\le q}$) remained above $0.95$
[Fig.~\ref{fig:results}(a)].  This
capacity increased from $B_0=33$ at $q=3$ to $B_0=450$ at $q=13$.  It
approximately followed $1/\pstar(q)$, and a finite-range fit gave
$B_0\propto q^{1.74}$ [Fig.~\ref{fig:results}(b)].  The fitted exponent
depends on the threshold and on the inference protocol: at a $0.90$
threshold, where the fixed-depth protocol of Table~\ref{tab:ladder} also
yields a complete ladder, the adaptive and fixed-depth exponents are $2.07$
and $1.69$, a gap confirmed on the four audit pools measured under both
protocols ($2.04$ against $1.69$), while the $0.95$ criterion is not
reliably attainable under
fixed depth (Supplemental Material).  The scale of these fits verifies the
mechanism expectation that capacity
tracks $1/\pstar(q)$, which grows as $q^{2}$ at large $q$.

For the matched classical best-of-$q$ control with the same table-based
policy, the corresponding capacity increased only from $11$ to $28$
question--answer pairs and fitted $B_0^{\rm cl}\propto q^{0.62}$.  The fitted
exponents differed by $1.12$, with a $95\%$ bootstrap interval of
$[0.98,1.31]$; the classical protocol involves no stopping choice, and a gap
of about one unit persists under the unified fixed-depth comparison at the
$0.90$ threshold (Supplemental Material).  Thus, over the measured range,
the coherent amplification budget increased the number of questions supported
by a fixed-size policy trained at that budget substantially faster than
classical repetition.  These are
finite-range empirical fits, not asymptotic complexity exponents.

\section{Conclusion and Outlook}
\label{sec:conclusion}

Slow thinking can be viewed as dynamics over reasoning trajectories.  An
autoregressive LLM defines a statistical distribution over possible action
sequences, but each inference run follows only one sampled sequence.  A quantum
AI model can instead evolve a coherent superposition of many sequences.  In a
general quantum readout, different histories can interfere when their
amplitudes are recombined; in the Grover realization studied here, this
recombination acts on the collective accepted and rejected components.
Quantum slow thinking is therefore a discrete path integral over reasoning
trajectories: the reasoning horizon sets the space of histories, while
interference determines how strongly different answers are represented.

Here we make this idea trainable using Grover amplitude amplification.
Classical RLVR rewards sampled trajectories that reach the correct answer.
Once one strategy succeeds more often, training repeatedly reinforces it and
may gradually suppress competing strategies.  With a finite number of Grover
rounds, by contrast, the highest final success can occur while the success
probability before interference is still below one.  Training therefore need not force every
problem to be solved almost certainly before the quantum evolution begins.  In
our models, this altered training pressure preserves multiple successful routes
while increasing the probability of obtaining the correct answer.  The
essential mechanism is not the use of quantum parameters alone: the same
quantum models undergo policy collapse when trained with the classical
single-attempt objective.  It is the joint design of training and quantum interference that
mitigates policy collapse.  Notably, the gain requires only this coarsest,
ensemble-level interference: no phase relation between individual reasoning
paths is used anywhere in the construction.

We demonstrate this mechanism on a $2\times3$ sliding-puzzle benchmark, in
which the target configuration is known but the sequence of moves leading to
it must be discovered.  Both the table-based QuCoNet and the
parameter-sharing neural quantum AI model retain branching ensembles of
reasoning paths.  At four Grover rounds, corresponding to nine applications of
the reasoning circuit, they achieve $3.2$ and $3.9$ times the
held-out accuracy of the strongest matched classical controls, respectively,
and retain $75$--$91\%$ of the held-out accuracy of the untrained
diversity-preserving reference, which none of the trained policies surpasses
on the
held-out questions of this solution-dense benchmark.  If $q$
denotes the number of applications of the reasoning circuit, the number of
question--answer pairs supported by a policy of fixed size, trained at each
budget, grows approximately as
$q^{1.74}$ under coherent quantum inference over the tested range, compared
with $q^{0.62}$ under classical repetition.  These results establish quantum
slow thinking as a trainable inference paradigm: the interior target set by
Grover inference preserves exploratory diversity, and the interference
converts interior success probabilities into verified performance.

The known-target benchmark allows every completed trajectory to be checked by
an exact verifier.  In generic question answering, however, the correct answer
is not supplied in advance.  A learned verifier or value model would then estimate
the correctness of candidate answers and
coherently convert this
estimate into a phase, without measuring or destroying the superposition of
reasoning paths.  Learning this evaluator jointly with the reasoning policy is
a direct route from the present construction toward general quantum
reasoning.  On the readout side, general non-diagonal measurements and
learned path-sensitive verifiers would engage interference between
individual reasoning paths, a resource the present two-sector mechanism
leaves untouched and a natural place to look for further advantages.  A physical realization will additionally require coherent memory,
reversible implementation of the learned evaluator, and control of noise and
of
fault-tolerance costs.

Classical slow thinking converts additional inference time into more numerous
or longer
sampled trajectories.  Quantum slow thinking converts the same resource into
longer coherent evolution over a superposition of reasoning histories.  Because
these histories remain coherent until their amplitudes are recombined,
exploration and selection become parts of the same quantum dynamics.  On the
benchmark studied here, this two-sector interference architecture enables
quantum slow thinking to maintain high verified accuracy without collapsing the
diversity of reasoning paths, while substantially outperforming matched
classical slow-thinking baselines.  Quantum-AI
architectures with richer interference may yield further advantages, either by
preserving still more trajectory diversity or by improving performance along
other dimensions; determining what these architectures can achieve remains an
open direction for future work.

\begin{acknowledgments}
X.C. and K.C. are supported by the Strategic Priority Research Program of
the Chinese Academy of Sciences under Grant No.~XDB1680102, the National
Key Research and Development Program of China under Grant
No.~2024YFA1408604, and the National Natural Science Foundation of China
under Grants No.~12474245 and No.~12447103.  X.C. is also supported by the
China Postdoctoral Science Foundation under Grant No.~2026M793770.
X.-H.D. thanks the
Shenzhen Science and Technology Program (KQTD20200820113010023) and the
Innovation Program for Quantum Science and Technology (2024ZD0300400).
\end{acknowledgments}

\bibliography{quantum_slow_reasoning}

\clearpage
\onecolumngrid
\begin{center}
  {\large\bfseries Supplemental Material for\\[0.35em]
  ``Do Quantum AIs Dream in Paths? Path-Integral Slow Thinking through
  Grover Interference''}
\end{center}
\vspace{0.5em}
\begin{quote}
This Supplemental Material derives the trajectory formulation and the
Grover-induced interior training target, proves monotonicity for
terminal-outcome-only classical sampling, and reports the complete task construction, model and
circuit definitions, control families, validation tests, scaling analysis,
and reproducibility information underlying the main text.
\end{quote}
\vspace{0.5em}

\setcounter{section}{0}
\setcounter{figure}{0}
\setcounter{table}{0}
\setcounter{equation}{0}
\setcounter{secnumdepth}{2}
\renewcommand{\thesection}{S\Roman{section}}
\renewcommand{\thefigure}{S\arabic{figure}}
\renewcommand{\thetable}{S\arabic{table}}
\renewcommand{\theequation}{S\arabic{equation}}
\renewcommand{\theHsection}{sm.\arabic{section}}
\renewcommand{\theHsubsection}{sm.\arabic{section}.\arabic{subsection}}
\renewcommand{\theHfigure}{sm.\arabic{figure}}
\renewcommand{\theHtable}{sm.\arabic{table}}
\renewcommand{\theHequation}{sm.\arabic{equation}}

\section{Classical trajectory measures and coherent path sums}
\label{sm:path-integral}

Let $x$ denote an input problem and
$\gamma=(z_1,\ldots,z_T)$ a complete sequence of intermediate reasoning
states or actions, where $T$ is the trajectory length.  Let $\theta$ denote
the trainable model parameters.  An autoregressive policy assigns
\begin{equation}
 P_{\theta}[\gamma|x]
 =\prod_{t=1}^{T}\pi_{\theta}(z_t|x,z_{<t})
 =e^{-I_{\mathrm{LLM}}[\gamma|x]} .
 \label{eq:sm-classical-path}
\end{equation}
Here $\pi_\theta(z_t|x,z_{<t})$ is the normalized conditional probability of
the next state, $z_{<t}=(z_1,\ldots,z_{t-1})$ is the complete prefix, and
$I_{\mathrm{LLM}}[\gamma|x]=-\log P_{\theta}[\gamma|x]$ is the trajectory
surprisal.  No Markov assumption is made: every factor may depend on the complete prefix.
The probability of an answer $y$ is a statistical sum over the
trajectories that produce it, and one inference run samples a realized
trajectory from this distribution.

A quantum reasoning circuit assigns amplitudes rather than probabilities,
\begin{equation}
 A_{\mathrm Q}[\gamma|x]
 =\prod_{t=1}^{T}\alpha_{\theta}(z_t|x,z_{<t})
 =|A_{\mathrm Q}[\gamma|x]|
   e^{iS_{\mathrm Q}[\gamma|x]/\hbar} ,
 \label{eq:sm-quantum-path}
\end{equation}
where $\alpha_\theta$ is a complex conditional transition amplitude,
$S_{\mathrm Q}$ denotes the accumulated effective phase action, and $\hbar$
fixes the phase unit and may be set to one; $S_{\mathrm Q}$ is defined modulo
$2\pi\hbar$.  We denote the corresponding state-preparation circuit
by $\mathcal A$.  Let $|\gamma\rangle$ denote the basis state containing
the complete trajectory record, and let $M_y$ be the positive measurement
operator, or POVM element, associated with outcome $y$.  The outcome probability is the double path sum
\begin{equation}
 P_{\mathrm Q}(y|x)=
 \sum_{\gamma,\gamma'} A_{\mathrm Q}[\gamma|x]
 A_{\mathrm Q}^{*}[\gamma'|x]
 \langle\gamma'|M_y|\gamma\rangle .
 \label{eq:sm-history-readout}
\end{equation}
The diagonal terms define a classical mixture of histories; the off-diagonal
terms encode coherent recombination.  Equations~\eqref{eq:sm-quantum-path}
and \eqref{eq:sm-history-readout} give the discrete path-integral
representation used in the main text.

For a general $M_y$, off-diagonal matrix elements allow different histories
to interfere when the measurement coherently recombines them.  In the Grover
realizations studied here, fresh which-path registers keep individual action
sequences orthogonal.  The verifier applies a common phase to the collective
accepted sector; by itself, this phase operation changes no probability.  The
state-preparation reflection, implemented with $\mathcal A^{\dagger}$ and
$\mathcal A$, converts the phase contrast between the accepted and rejected
components into a coherent rotation of their weights.  No pairwise
interference between individual path records contributes to the success
probability or to the path-resolved observables studied here, nor is such
pairwise interference required for amplitude amplification.

\section{Tasks, models, and training protocol}
\label{sm:protocol}

\subsection{Reasoning tasks and data splits}

The primary task is goal-directed navigation on the configuration graph of a
$2\times3$ sliding puzzle.  Under single slides the puzzle has $6!/2=360$
reachable configurations and a hole-position-dependent branching number.  To
obtain a regular graph we restrict the hole to the two middle-column cells and
use three self-inverse compound moves: a vertical slide $V$ exchanging the
hole between the two middle cells, and two three-slide cycles $L$ and $R$ that
carry the hole to the other middle cell around the left or right pair of
corners.  Slide moves preserve the parity of tile inversions~\cite{wilson1974puzzles}, so of the
$2\times5!=240$ middle-column arrangements exactly the $120$ sharing the
goal's parity are reachable and form a closed component: $N=120$ nodes of
degree $K=3$, each generator being its own inverse.  A question is an ordered pair
$(Q,A)$ at shortest-path distance $3$--$6$; a trajectory succeeds on its first
arrival at $A$ within $M=8$ moves.  For each question-pool seed we draw $192$ pairs.
The first $B$ pairs form the training set and the last $64$ form the
held-out set, shared by all model families and all
values of $B$ for that seed; the two sets contain no common pair (zero
collisions across all question-pool seeds and training-set sizes).  Unless noted
otherwise, reported tabular points pool $32$ question-pool seeds while fixing
the optimization seed at $42$.  A pool--optimizer-seed cross-audit at $B=32$
(Grover $q=3$, Grover $q=9$, and best-of-four objectives on pools $1$--$4$,
each retrained with three additional optimization seeds) found within-pool
optimizer-seed standard deviations of $0.010$--$0.026$ across objectives and
splits, comparable to the matched pool-to-pool values of $0.008$--$0.028$
(both ranges exclude the saturated Grover $q=9$ training cells, where both
spreads vanish),
so the question-pool ensembles are not hiding optimization scatter.  Every
Grover $q=9$ run reached training accuracy $1.000$ under the adaptive rule,
and no Grover-trained run of the audit placed any training question at the
deterministic boundary ($p_i>0.98$) at these budgets, whereas the
best-of-four control placed up to two of $32$ questions there per pool.  The $q=7,9$ scans away from
$B=32$, the random-regular replication, and the large-size scans use eight
question-pool seeds.  The large-$B$ capacity scan of Sec.~\ref{sm:scaling} uses the same
construction with $16$ larger pools: $768$ pairs per seed, extended
to $1280$ pairs for $B>704$, with the first $704$ training pairs and the
held-out set identical across the two pools so that the two curves stitch
together.
For $q=3,5,7,9$, the coherent grid is
$B\in\{8,16,32,64,96,128,192,256,384,512,704\}$; the $q=11$ scan adds
$B=768,896,1024$, and the $q=13$ scan additionally includes $B=1216$.
The matched classical capacity control uses those same $16$ $768$-pair pools,
the same ordered training prefixes, and
$B\in\{1,2,4,6,8,12,16,24,32,40,48,64,96,128\}$.
Here and below, $q$ denotes the total number of forward or inverse applications
of the reasoning circuit.  Thus $q=2r+1$ for $r$ Grover rounds.
Throughout the Supplemental Material, accuracy denotes the mean final success
probability over the specified set of question--answer pairs.

As a topology control, we generate uniform random $3$-regular graphs with the
same $N$, horizon, distance filter, and question protocol.  A size scan uses
$N\in\{120,240,480\}$ at $B=32$.  These are not intended as language tasks;
they isolate the relation among policy support, a verifiable trajectory
predicate, and the inference map.

\subsection{Classical and quantum walkers}

The classical table policy, CoNet, has unconstrained real parameters, or
logits, $\ell\in\mathbb R^{N\times K}$.  At node $x$ it selects edge $k$ with
probability
$\pi_x(k)=e^{\ell_{xk}}/\sum_{k'}e^{\ell_{xk'}}$.  The sampled baseline
uses REINFORCE~\cite{williams1992reinforce}, binary rewards, and a
per-question mean baseline.  Exact-gradient controls differentiate through the
absorbing Markov dynamic program described in Sec.~\ref{sm:exact}.  A
repeated-sampling (best-of-$k$) control, which succeeds if any of $k$
independent completed trajectories passes the
verifier~\cite{chow2024inference}, minimizes
\begin{equation}
  \mathcal L_{\mathrm{cl},k}
  =-\sum_i\left[1-(1-p_i)^k\right].
  \label{eq:bestof-loss}
\end{equation}

The autoregressive (sequentially conditioned) quantum table policy, QuCoNet,
has a coin
$C_x=\exp(iH_x)$ at every node,
where $H_x$ is a real symmetric $3\times3$ matrix.  It therefore has six real
parameters per node and $720$ parameters in total.  At step $s$, a
question-dependent basis channel $c_s\in\{0,\ldots,K-1\}$ enters the coin.
The length-$M$ channel sequence is a deterministic code for $(Q,A)$, formed by
interleaving the base-$K$ digits of the start and target node labels and
repeating or truncating the pattern to length $M$.  At $M=8$ the
truncation drops one base-$3$ digit of each label, so the code is not
injective: across the $32$ mainline pools, $18$ training questions share a
channel sequence with another training question of the same pool, and $66$
of the $2{,}048$ held-out questions ($3.2\%$) share one with a training
question.  The node labels themselves carry no information the model can
exploit: retraining the Grover $q=3$ objective on the eight mainline pools
after two independent uniformly random node relabelings (graph, questions,
and answers relabeled together; optimization seed fixed) reproduced the
native results: the relabeled-minus-native differences are
$-0.006\pm0.032$ in training accuracy and $+0.007\pm0.023$ held-out
(mean $\pm$ sample standard deviation over the $16$ runs), both consistent
with zero.  If $x_k$ denotes the
neighbor reached from node $x$ through local edge $k$, the training-time
transition probability is
\begin{equation}
  P(x\rightarrow x_k\mid c_s)=|C_x[c_s,k]|^2 .
  \label{eq:coin-probability}
\end{equation}
Fresh path-record registers make different action histories orthogonal.  Hence the
position marginal during training is exactly the classical Markov chain of
Eq.~\eqref{eq:coin-probability}, even though the same gates define a coherent
unitary at inference.  The classical single-attempt objective trains the
success probability of one completed attempt, whereas the Grover-based
objective trains the success after coherent inference.  Their losses are
$-\sum_i p_i$ and $-\sum_iG_{\le q}(p_i)$
(Sec.~\ref{sm:objective}), respectively.  Both use
exact gradients.

The two models are matched in graph interface, horizon, training-set size, and
optimization protocol, not in raw parameter count ($360$ stochastic logits
versus $720$ real coin parameters).  This mismatch cannot explain the central
comparison: the monotonicity result in Sec.~\ref{sm:proof} applies to any policy
architecture when classical procedures access only the terminal binary
verifier outcome of each independent completed sample,
while the exact-gradient classical controls remove
estimator noise altogether.

\section{Exact evaluation and path support}
\label{sm:exact}

For a question $(Q,A)$, depth-first enumeration visits all action sequences of
length at most $M$.  A branch terminates and contributes its probability when
it first reaches $A$; branches below probability $10^{-15}$ are pruned.  An
equivalent memory-light computation propagates a position occupancy vector,
zeros the target outflow at every step, and accumulates its inflow.  Enumeration
and this absorbing dynamic program agree to $10^{-9}$ on random checkpoints;
the remaining precision is set by the stored \texttt{float32} parameters,
which can overshoot unit probability by up to $\sim4\times10^{-6}$ on
saturated questions; all read-out maps clip $p$ to $[0,1]$ before use.

\begin{figure}[t]
  \includegraphics[width=0.90\textwidth]{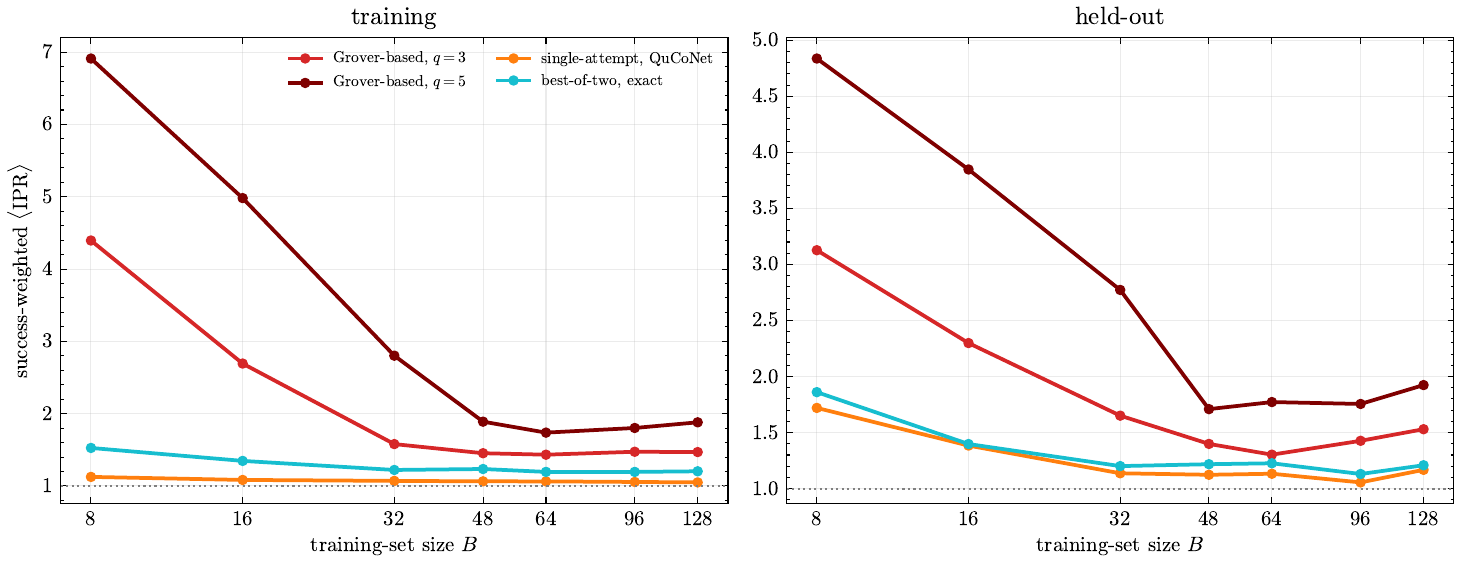}
  \caption{Effective successful-path number on training (left) and held-out
  (right) questions.  Grover-based training retains multiple effective
  paths, whereas single-attempt and best-of-two objectives approach the single-path
  reference (dotted).  The inverse participation ratio is weighted by each
  question's success probability.}
  \label{fig:sm-ipr}
\end{figure}

Enumeration also gives individual successful-path probabilities $w_j$.  With
$\widetilde w_j=w_j/\sum_jw_j$, we quantify the effective number of routes~\cite{belldean1970,wegner1980ipr} by
\begin{equation}
  \mathrm{IPR}=\frac{1}{\sum_j\widetilde w_j^2}.
  \label{eq:ipr}
\end{equation}
Batch means are weighted by question success probability,
$\sum_i p_i\mathrm{IPR}_i/\sum_i p_i$, so dead questions do not inflate or
dilute the measure.  Figure~\ref{fig:sm-ipr} shows that Grover-based
training retains several effective routes, whereas the effective path number
under single-attempt and repeated-sampling training approaches one.  The figure shows $q=3,5$; the retained ensemble grows
with the intended number of circuit applications, with accuracy-weighted
$\langle\mathrm{IPR}\rangle=1.6,\,2.8,\,4.2,\,5.7$ on training questions at
$B=32$ for $q=3,5,7,9$ (held-out values within $0.4$ of these).  The
weighted means summarize broad per-question distributions: pooled over the
eight shared pools at $B=32$, the training IPR at $q=3$ has median $1.19$
with interquartile range $[1.00,1.96]$ ($23\%$ of questions above $2$), and
at $q=9$ median $5.50$ with interquartile range $[3.45,8.35]$ ($92\%$ above
$2$, maximum $16.2$).  The inverse
participation ratio (IPR) is a path-count diagnostic; it is not evidence for
interference among those routes.  It is also strictly invariant under the
amplification itself: $r$ Grover rounds multiply the entire accepted
component by a common factor, so the conditional weight of every accepted
path, $\widetilde w_j=w_j/\sum_j w_j$, is exactly unchanged, and the
training loss, a function of $p_i$ alone, is constant on surfaces of fixed
$p_i$ regardless of how the success weight distributes over routes.  Any
change in IPR is therefore a property of the optimization dynamics toward
the interior target, not of the Grover map, consistent with the two-sided
control of Sec.~\ref{sm:controls}.

\begin{figure}[t]
  \includegraphics[width=\linewidth]{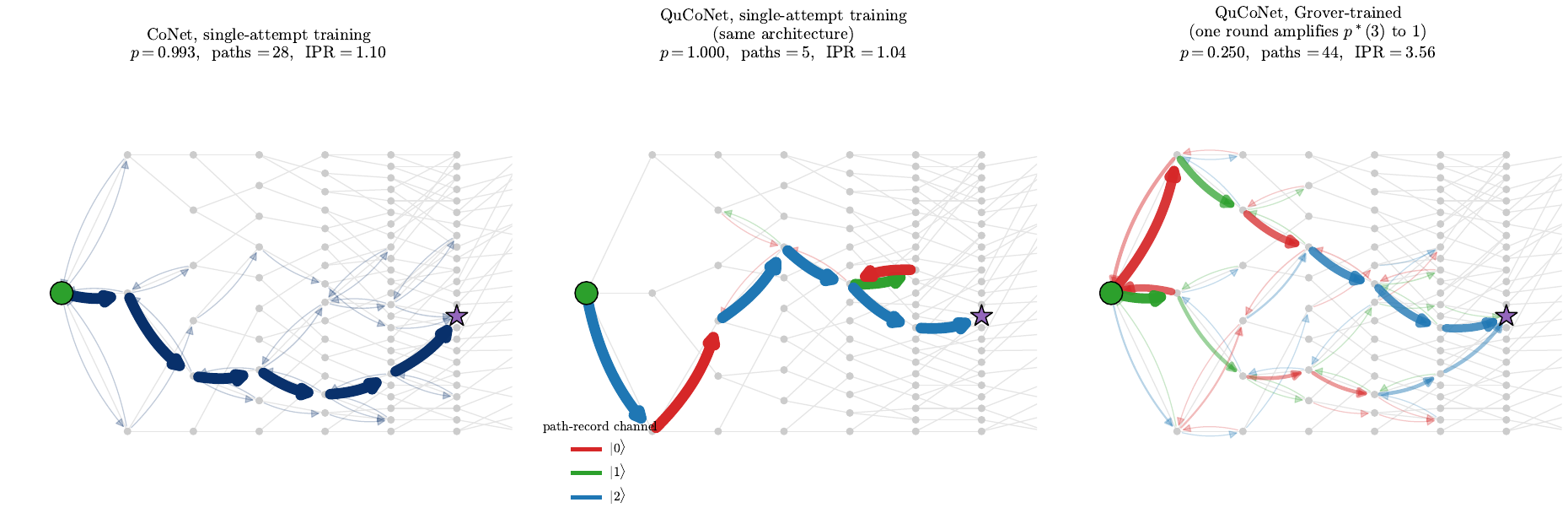}
  \caption{Success-probability flow for one training question
  ($B=32$, question-pool seed~$1$;
  start, green circle; target, purple star), with nodes ordered left to right
  by graph distance from the start and each edge's width giving its share of
  the question's total success probability.  Left: the single-attempt-trained CoNet policy
  collapses onto a single route ($p=0.993$, $\mathrm{IPR}=1.10$).  Middle: the
  QuCoNet policy of identical architecture, trained with the single-attempt
  objective, collapses in the
  same way ($p=1.000$, $\mathrm{IPR}=1.04$), so the collapse comes from the
  objective, not the substrate.  Right: the same QuCoNet policy trained for one
  Grover round holds the question at $\pstar(3)=1/4$ with a branching ensemble
  ($p=0.250$, $\mathrm{IPR}=3.56$), colored by orthogonal path-record channel.
  The inverse participation ratio counts effective routes and is not evidence of
  interference among them.}
  \label{fig:sm-flows}
\end{figure}

Figure~\ref{fig:sm-flows} makes the same point on a single question by drawing
where the trained success probability flows.  Single-attempt-trained CoNet and
single-attempt-trained QuCoNet both collapse onto one
route, so the collapse is a property of the objective rather than the substrate;
only Grover-based training holds the question at $\pstar(3)$ with a branching
ensemble.  Panels (b),(c) of the main-text opening figure reproduce the left
(classical single-attempt) and right (Grover-trained) panels.

All matched-cost comparisons count applications of the reasoning circuit
$\mathcal A$ or its inverse.  Preparing the state costs one application, and
each Grover round adds $\mathcal A^\dagger$ and $\mathcal A$, so $r$ rounds
use $q=2r+1$ circuit applications.  The coherent procedure is therefore
compared with $q$ independent classical attempts.  This cost model excludes
the $r$ coherent verifier reflections and also assigns no cost to the
classical verifier outcomes.  The comparison point moves in either direction
when these choices vary.  If a coherent reflection costs $\lambda_{\rm v}$
circuit-application units and a classical verification costs
$\lambda_{\rm c}$ units, the budget that funds $r$ Grover rounds funds
$k=[(2r+1)+\lambda_{\rm v}r]/(1+\lambda_{\rm c})$ classical attempts.  Our
convention $\lambda_{\rm v}=\lambda_{\rm c}=0$ gives $k=q=2r+1$.  Charging
only the quantum side ($\lambda_{\rm v}=1$, $\lambda_{\rm c}=0$) raises the
classical allowance at $r=4$ from $9$ to $13$ attempts; recomputing the
best-of-$k$ envelope of all $27$ controls at $k=13$ raises the strongest
classical held-out mean from $0.126$ to $0.156$, so the main-text ratios
$3.2$ and $3.9$ become $2.6$ and $3.2$ with the ordering unchanged.
Charging both sides equally ($\lambda_{\rm v}=\lambda_{\rm c}=1$) instead
lowers the classical allowance to $k=(3r+1)/2\approx1.5r$, fewer than in our
comparison.  Coherent synthesis of the verifier reflection is an additional
physical resource and is discussed as a limitation in
Sec.~\ref{sm:limitations}.

\section{Classical sample-access monotonicity}
\label{sm:proof}

The main-text monotonicity observation includes two assumptions that are essential for its
scope.  First, the verifier is sound: an accepted completed trajectory is a
valid solution.  Second, the procedure is success maximizing: once at least
one accepted solution has been observed, it retains or returns an accepted
solution rather than intentionally converting that event into failure.
These assumptions exclude perverse rules such as
``reject whenever a success is observed,'' which are nonmonotone but useless
for solving the task.

\emph{Proof.}  Let $p<p'$.  Couple attempt $t$ at the two single-attempt success
probabilities by drawing the same $u_t\sim\operatorname{Uniform}[0,1]$ and
declaring the outcome successful at $p$ if $u_t<p$, and at $p'$ if $u_t<p'$.
Realize the procedure's private randomness as a single sequence drawn once and
shared between the two worlds, and let $\tau$ be the first attempt at which
the coupled outcomes differ, that is, the first $t$ with $u_t\in[p,p')$.
Before $\tau$ the two worlds hold identical verifier histories and private
randomness, hence take identical adaptive actions: when to stop, whether to
draw again, which verified solution to return.  If $\tau$ never
occurs before the procedure stops, the two runs coincide and either both
succeed or both fail.  If it does occur, the run at $p'$ has just observed a
verified success, and a sound, success-maximizing rule must therefore succeed.
The run at $p$ may still succeed later, but cannot outperform a success.  On
every realization the run at $p'$ therefore does at least as well, and its
success probability is nondecreasing in $p$. \hfill$\square$

The statement is deliberately an access theorem, not a theorem that every
conceivable classical algorithm is monotone in every internal policy
parameter.  It covers procedures that act on independent completed samples
only through their verifier verdicts; a rule that keys on trajectory
content can correlate with the identity of the policy and evade the
coupling.  It does not cover a
procedure with coherent access to amplitudes, a procedure that changes the
underlying generator between attempts using extra semantic information, or a
manually imposed nonmonotone utility.  Tree search may allocate samples much
better than naive independent repeated sampling; so long as success is obtained
by observing
verified classical trajectories, however, increasing the probability of an
otherwise identical successful draw cannot hurt.

\section{Amplification objective and stopping rule}
\label{sm:objective}

\begin{figure}[t]
  \includegraphics[width=\linewidth]{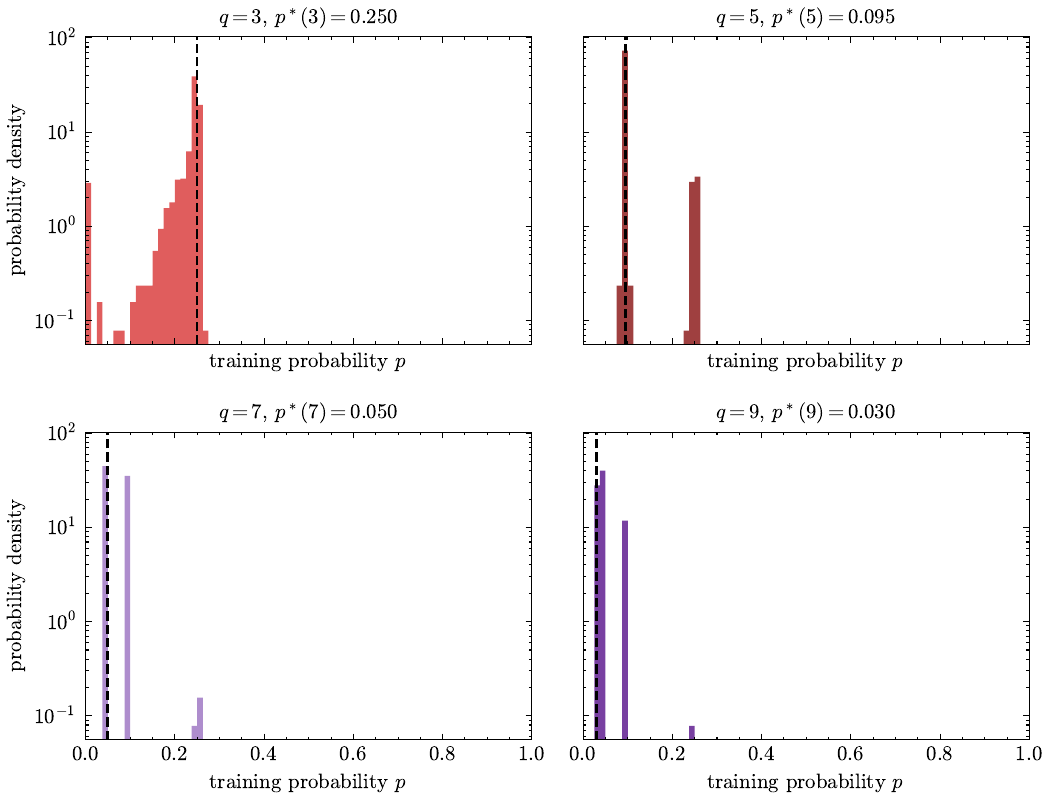}
  \caption{Exact training-set distributions of the per-question success
  probability after Grover-based training for $q=3,5,7,9$ ($B=32$,
  $32$ question-pool seeds each).  Dashed lines mark the predicted interior optimum
  $\pstar(q)=\sin^2[\pi/(2q)]$ of Eq.~\eqref{eq:ridges}; panel titles give
  the intended number of circuit applications and its interior
  optimum.  The distribution concentrates
  at $\pstar(q)$ in every panel; smaller
  peaks at $q\ge5$ lie near higher-order maxima requiring fewer than $q$
  circuit applications.}
  \label{fig:sm-nscan}
\end{figure}

Write the prepared state as
\begin{equation}
  |\psi\rangle=\mathcal A|0\rangle
  =\sin\vartheta\,|\psi_{\mathrm{acc}}\rangle
  +\cos\vartheta\,|\psi_{\mathrm{rej}}\rangle,
  \qquad \sin^2\vartheta=p,\quad 0\le\vartheta\le\frac{\pi}{2},
  \label{eq:good-bad}
\end{equation}
where $|\psi_{\mathrm{acc}}\rangle$ and $|\psi_{\mathrm{rej}}\rangle$ are the normalized projections
onto the verifier-accepted and rejected subspaces.  Standard amplitude
amplification~\cite{grover1996search,brassard2002amplitude} gives
$G_q(p)=\sin^2(q\vartheta)$ for an odd number $q$ of circuit applications.  To avoid
deliberate overshoot in the training objective, we use
\begin{align}
  G_{\le q}(p)&=G_{\widehat q(p;q)}(p),\label{eq:Gleq-sm}\\
  \widehat q(p;q)&=2\min\!\left[\frac{q-1}{2},r^{*}(p)\right]+1,\nonumber\\
  r^{*}(p)&=\max\!\left[0,\operatorname{round}
  \left(\frac{\pi}{4\arcsin\sqrt p}-\frac12\right)\right].\nonumber
\end{align}
Here $r^{*}(p)$ is the optimal round count of quantum
search~\cite{boyer1998tight}; for $p=0$ we set $r^{*}(0)=0$, so
$G_{\le q}(0)=0$.  The interior maxima
\begin{equation}
  \pstar(q')=\sin^2\!\frac{\pi}{2q'},\qquad q'=3,5,\ldots,q,
  \label{eq:ridges}
\end{equation}
are the base probabilities amplified exactly to unit success.  Among these,
$\pstar(q)$ is the smallest and is the maximum reached from the initialized
small-$p$ regime.  For
$q\ge5$, smaller secondary peaks in the histograms of Fig.~\ref{fig:sm-nscan}
lie near higher-order maxima from which fewer rounds suffice.

Equation~\eqref{eq:Gleq-sm} is an objective-level optimal-stop convention; it
nominally uses $p$.  The main held-out comparison does not rely on it:
every held-out accuracy in the main text is reported
under a fixed-depth protocol that always uses all $q$ circuit applications and never
consults $p$.  The convention is nearly immaterial there: at $B=32$ the
fixed-depth sequence $0.042$, $0.147$, $0.272$, $0.407$ for $q=3,5,7,9$ lies within
$0.012$ of the $G_{\le q}$ values $0.043$, $0.149$, $0.278$, $0.418$, and
the difference stays below $0.026$ over the entire training-set-size
grid.  The held-out per-question probabilities lie almost entirely below the lowest maximum, where the
two rules coincide.  The convention matters only for the near-unity
training-set accuracies at $q\ge5$, where the smaller peaks at higher-order
maxima overshoot the deepest fixed-depth protocol: the $G_{\le q}$ training
accuracies at $B=32$, which are $0.953$, $1.000$, $1.000$, $1.000$ for
$q=3,5,7,9$ (question-pool standard deviation $0.021$ at $q=3$ and below
$10^{-3}$ beyond),
fall to $0.953$, $0.940$, $0.846$,
$0.772$ when every question receives all $q$ circuit applications.  A
fixed-depth implementation can
recover the training-side values with a fixed-depth $G_q$ training objective,
fixed-point amplitude amplification~\cite{yoder2014fixedpoint}, or an
amplitude-estimation stopping
rule~\cite{brassard2002amplitude}.  Training-side accuracies in the main text use the $G_{\le q}$
convention of Eq.~\eqref{eq:Gleq-sm}, which is the objective being optimized;
held-out numbers use the fixed-depth protocol throughout.

Training at larger $q$ also produces a better held-out policy at every tested
inference depth, rather than merely applying deeper inference to the same
residual probability.  Table~\ref{tab:cross} evaluates each model trained at
$q$ for every inference depth $q'$: every column increases down the rows.  The
$q{=}9$-trained model has the highest mean even at a single round, and the
$q{=}3$-trained model reaches only $0.111$ at nine circuit applications, little
more than a quarter of the
$q{=}9$-trained value.

\begin{table}[t]
\caption{Cross-depth evaluation on held-out questions at $B=32$: rows are
training objectives and columns are the fixed inference depths, using all
$q'$ circuit applications;
the optimal-stop rule raises the diagonal by at most $0.012$ and
the off-diagonal cells, which overshoot more often, by up to $0.03$.  The
  $q=5$--$9$ entries at $q'{=}3$ lie within one seed standard
deviation of one another.}
\label{tab:cross}
\begin{ruledtabular}
\begin{tabular}{lcccc}
 & $q'{=}3$ & $q'{=}5$ & $q'{=}7$ & $q'{=}9$\\
\midrule
trained at $q=3$ & 0.042 & 0.067 & 0.091 & 0.111\\
trained at $q=5$ & 0.069 & 0.147 & 0.215 & 0.263\\
trained at $q=7$ & 0.076 & 0.176 & 0.272 & 0.349\\
trained at $q=9$ & 0.079 & 0.192 & 0.309 & 0.407
\end{tabular}
\end{ruledtabular}
\end{table}

At $B=32$, the results across inference depths are summarized in Table~I of the main text
of the main
text; its classical row is the best-performing control at each $q$ in the audit of
Sec.~\ref{sm:controls} (Table~\ref{tab:controls}), granted the same $q$
verified samples.  On this
puzzle the one-round tabular system is not yet better on held-out questions;
its decisive held-out separation begins at larger $q$.  This is why the
main text reports the entire sequence rather than presenting a universal
one-round held-out advantage.

\begin{figure}[t]
  \includegraphics[width=\linewidth]{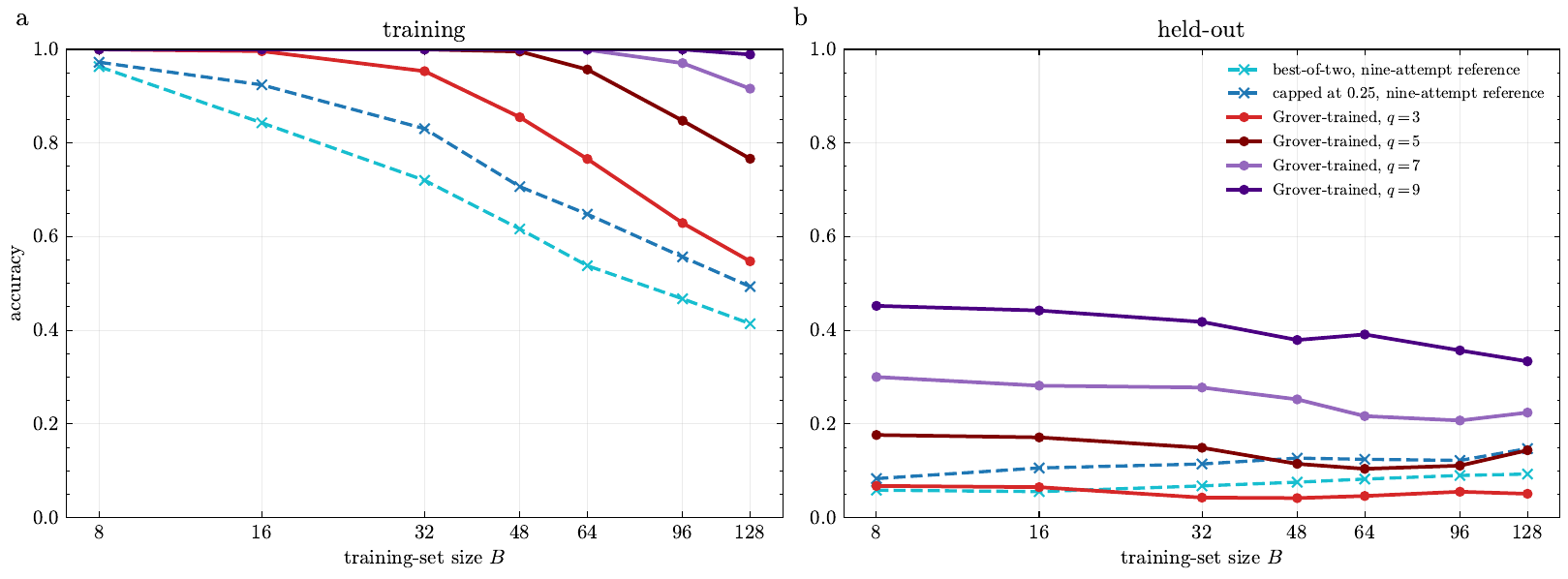}
  \caption{\textbf{Performance at increasing coherent inference depth.}  The
  Grover-trained model is evaluated at its matched number $q$ of circuit
  applications under the $G_{\le q}$ rule, which stays within $0.026$
  of the fixed-depth protocol on held-out questions.  The exact best-of-two
  and capped $c{=}0.25$ walkers are evaluated with nine attempts on training
  (a) and held-out (b) questions across training-set size $B$ on the sliding
  puzzle; the best-performing control in Table~\ref{tab:controls} at each $q$
  is slightly stronger.  Grover-based objectives at larger $q$ improve
  throughout and the ordering with $q$ stays monotone, while these fixed-depth
  classical references do not separate.  The held-out panel (b) is the full-$B$ view of the
  $B{=}32$ ladder in Table~I of the main text.  The dedicated $16$-pool scan
  used for the main-text capacity fit is reported in Sec.~\ref{sm:scaling}.}
  \label{fig:sm-ladder}
\end{figure}

\section{Coherent inference circuit}
\label{sm:circuit}

\begin{figure}[t]
  \includegraphics[width=\linewidth]{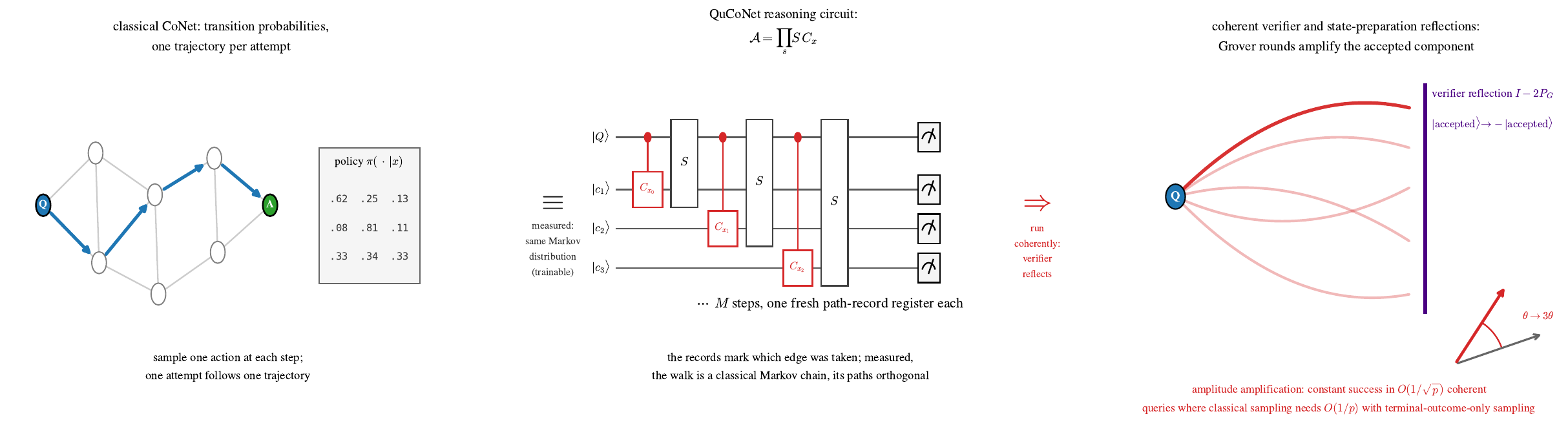}
  \caption{Classical sampling and coherent inference use the same local walk.
  Left: CoNet samples one action from $\pi(\cdot\mid x)$ at each node and
  returns one trajectory per attempt.  Middle: QuCoNet applies the local coin
  operator $C_x$ and a reversible shift $S$ at each step.  The shift moves to
  the neighboring node while recording the action in a fresh which-path
  register, giving $\mathcal A=\prod_s S C_x$.  Measuring these records
  reproduces the CoNet trajectory distribution.  Right: when the same
  reasoning circuit is kept coherent, the verifier reflection
  $I-2P_{\mathrm{acc}}$ gives
  the accepted and rejected components opposite phases.  The reflection about
  the prepared state converts this phase difference into amplitude
  amplification.  After $r$ rounds the accepted-sector angle is
  $(2r+1)\vartheta$.  No pairwise interference between individual path records is
  required.}
  \label{fig:sm-model}
\end{figure}

Figure~\ref{fig:sm-model} summarizes the construction of this section: the walk
that trains as a classical Markov chain is, run coherently, a
reasoning circuit whose accepted and rejected components are rotated by the
Grover reflections.

\subsection{State preparation and reversible shift}

Let $\mathcal H_{\mathrm{pos}}$ be the position space spanned by the $N$ graph
nodes and let $\mathcal H_{c_s}$ be the $K$-dimensional record for the action
taken at step $s$.  The coherent Hilbert space is
\begin{equation}
  \mathcal H=\mathcal H_{\mathrm{pos}}\otimes
  \mathcal H_{c_1}\otimes\cdots\otimes\mathcal H_{c_M},
  \qquad \dim\mathcal H=N K^M=787{,}320 .
  \label{eq:hilbert}
\end{equation}
It embeds in $\lceil\log_2N\rceil+M\lceil\log_2K\rceil=23$ qubits, counting
only the logical position and path-record registers; verifier workspace,
comparator, and readout ancillas are additional
(Sec.~\ref{sm:limitations}).  Unused basis states are padded by identity
action.  Step $s$ applies the current
node's coin operator to register $c_s$ and then a flip-flop shift, a reversible
operation that moves to a neighboring node while recording the traversed action,
\begin{equation}
  |x,k\rangle\longmapsto
  |\operatorname{nbr}(x,k),\operatorname{rev}(x,k)\rangle .
  \label{eq:shift}
\end{equation}
Here $\operatorname{nbr}(x,k)$ is the neighboring node reached through local
edge $k$, and $\operatorname{rev}(x,k)$ is the label of the same edge at that
neighbor.  This reverse-edge label records the action needed to undo the move.
The resulting map is a permutation of the joint position-register
basis and hence a unitary; a neighbor column by itself need not be a
permutation.  Eight steps form $\mathcal A$.

\subsection{First-arrival oracle without measurement}

Absorbing classical evaluation stops on first arrival, whereas a unitary
circuit cannot stop by measurement.  Each final basis state nevertheless
contains a complete action record, and the reversible shifts decode it to a
unique path.  The accepted-subspace projector $P_{\mathrm{acc}}$ is therefore
diagonal: it accepts a
final basis state if and only if its decoded path first reaches the target at
some step $s\le M$.  A reversible implementation~\cite{bennett1973reversible} computes this Boolean
predicate into an ancilla, applies a phase, and uncomputes it.

The Grover iterate can be written
\begin{equation}
  \mathcal Q=
  \mathcal A(2|0\rangle\!\langle0|-I)\mathcal A^\dagger
  (I-2P_{\mathrm{acc}}) .
  \label{eq:iterate}
\end{equation}
Although $|\psi_{\mathrm{acc}}\rangle$ is a high-dimensional entangled
superposition of many trajectories, $\mathcal Q$ preserves the two-dimensional
plane spanned by $|\psi_{\mathrm{acc}}\rangle$ and
$|\psi_{\mathrm{rej}}\rangle$.  The standard rotation therefore depends only
on $p=\langle\psi|P_{\mathrm{acc}}|\psi\rangle$.  Orthogonal which-path records
do not obstruct the rotation, and no cancellation among individual accepted
paths is needed.

\subsection{Full-state verification}

We propagated every training and held-out question for the one-round-trained
model at question-pool seed 1 and $B=32$ through the full state space of
Eq.~\eqref{eq:hilbert}.  The circuit and enumeration are independent
implementations.  Table~\ref{tab:circuit} shows single-precision agreement.  The first
row is the central implementation check: the coherent, never-absorbed circuit
has exactly the same accepted-subspace weight as the absorbing-state calculation
used for training.  In Table~\ref{tab:circuit}, $p_i^{\rm enum}$ is the
independently enumerated success probability and
$\sin^2\vartheta_i=p_i$ for question $i$.

\begin{table}[t]
\caption{Full-Hilbert-space verification.  Maxima are over $32$ training and
$64$ held-out questions.  The after-round value in the last row is the mean
of $G_{3}(p_i)$ over questions, not $G_{3}$ of the mean.}
\label{tab:circuit}
\begin{ruledtabular}
\begin{tabular}{lc}
Check & value\\
\midrule
$\max_i|\langle\psi_i|P_{\mathrm{acc}}|\psi_i\rangle-p_i^{\rm enum}|$
  & $3.6\times10^{-7}$\\
$\max_i|\|P_{\mathrm{acc}}\mathcal Q|\psi_i\rangle\|^2
-\sin^2(3\vartheta_i)|$
  & $1.5\times10^{-5}$\\
$\max_i|\|P_{\mathrm{acc}}\mathcal Q^2|\psi_i\rangle\|^2
-\sin^2(5\vartheta_i)|$
  & $1.3\times10^{-5}$\\
Mean training $p$ / after one round & $0.243/0.970$
\end{tabular}
\end{ruledtabular}
\end{table}

\section{Classical controls and diagnostic questions}
\label{sm:controls}

The controls separate five possible explanations of the observed difference:
gradient-estimator noise, model architecture, loss shape, post-training
coherent inference, and the favorable uniform prior of the small graph.

\emph{Does estimator noise explain the difference?}  Differentiating
Eq.~\eqref{eq:bestof-loss} removes Monte Carlo gradient-estimation noise and implements the
inference-aware classical objective~\cite{chow2024inference} directly.  It closely reproduces the sampled
best-of-two result across $B$.  At $B=32$, the exact-gradient and sampled
models reach $0.657$ and $0.659$, respectively, when granted three attempts,
below the one-round system's $0.953$.  This comparison isolates the objective
from the estimator.

\emph{What is the strongest terminal-outcome-only baseline?}
Table~\ref{tab:controls} audits the 27 controls that access only the final
binary verifier outcome.  They include single-attempt, best-of-$k$ up to
$k=64$, capped, and
entropy-regularized objectives on the softmax walker, and the same classical
objectives trained on the coin architecture itself, which the which-path
structure makes a classical Markov model.  All matched-cost comparisons in
the main text use the best result from this family at each $q$ and data split.
At three circuit applications the strongest member on training questions is the
QuCoNet-architecture
best-of-four control ($0.729$); on held-out questions the strongest members
are the best-of-$k$ controls with large $k$ ($0.071$ to $0.126$ from three to
nine applications).  The family exhibits a trade-off between accuracy and
coverage.  Objectives that perform well at small $q$ concentrate probability
on fewer questions and generalize less well, while best-of-$64$ retains broader
coverage but cannot convert it into high accuracy at small $q$.  No single
control achieves both.  The single-attempt, best-of-two, capped, and
large-$k$ softmax families span the full $B$ grid; the entropy-regularized and
QuCoNet-architecture families exist at $B\in\{8,32,128\}$.  At each $B$, the
reported classical control is the best available member of this family.
In Table~\ref{tab:controls}, ``single-attempt, Adam'' denotes the earlier
$120$-epoch Adam run, whereas ``single-attempt, protocol-matched'' uses the
same $200$-epoch optimization and model-selection schedule as the core
Grover-trained tabular runs.

\begin{table}[t]
\caption{Control audit at $B=32$: best-of-$q$ accuracy of the 27 classical
objectives on training and held-out questions.  Results pool $32$ question-pool seeds for every family.
``QuCoNet arch.''\ rows train a classical objective using the same local coin
operators as QuCoNet.  The main text's ``strongest classical control'' is the
best result in this table at each $q$ and data split.}
\label{tab:controls}
\begin{ruledtabular}
\begin{tabular}{lcccccccc}
 & \multicolumn{4}{c}{training} & \multicolumn{4}{c}{held-out}\\
\cmidrule(lr){2-5}\cmidrule(lr){6-9}
objective & $q{=}3$ & $q{=}5$ & $q{=}7$ & $q{=}9$ & $q{=}3$ & $q{=}5$ & $q{=}7$ & $q{=}9$\\
\midrule
single-attempt, REINFORCE & 0.602 & 0.621 & 0.632 & 0.639 & 0.057 & 0.066 & 0.071 & 0.075\\
single-attempt, Adam & 0.537 & 0.545 & 0.551 & 0.555 & 0.049 & 0.052 & 0.055 & 0.057\\
single-attempt, protocol-matched & 0.533 & 0.540 & 0.544 & 0.547 & 0.045 & 0.048 & 0.050 & 0.052\\
single-attempt, exact gradient & 0.535 & 0.541 & 0.545 & 0.548 & 0.047 & 0.050 & 0.052 & 0.053\\
best-of-2, sampled & 0.659 & 0.701 & 0.716 & 0.723 & 0.053 & 0.060 & 0.065 & 0.068\\
best-of-2, exact & 0.657 & 0.699 & 0.714 & 0.720 & 0.054 & 0.060 & 0.065 & 0.068\\
best-of-4, exact & 0.681 & 0.755 & 0.784 & 0.796 & 0.062 & 0.074 & 0.081 & 0.086\\
best-of-8, exact & 0.664 & 0.780 & 0.831 & 0.856 & 0.070 & 0.087 & 0.099 & 0.107\\
best-of-16, exact & 0.601 & 0.750 & 0.829 & 0.873 & 0.071 & 0.091 & 0.105 & 0.116\\
best-of-32, exact & 0.486 & 0.655 & 0.762 & 0.832 & 0.065 & 0.090 & 0.109 & 0.124\\
best-of-64, exact & 0.352 & 0.508 & 0.623 & 0.710 & 0.058 & 0.085 & 0.107 & 0.126\\
capped $c{=}0.25$ & 0.607 & 0.736 & 0.799 & 0.831 & 0.070 & 0.090 & 0.104 & 0.114\\
capped $c{=}0.50$ & 0.656 & 0.714 & 0.732 & 0.740 & 0.061 & 0.072 & 0.078 & 0.083\\
capped $c{=}0.75$ & 0.628 & 0.660 & 0.675 & 0.684 & 0.052 & 0.059 & 0.064 & 0.068\\
single-attempt $+\lambda H$, $\lambda{=}0.01$ & 0.558 & 0.566 & 0.570 & 0.572 & 0.055 & 0.065 & 0.072 & 0.078\\
single-attempt $+\lambda H$, $\lambda{=}0.03$ & 0.558 & 0.565 & 0.568 & 0.570 & 0.060 & 0.072 & 0.080 & 0.086\\
single-attempt $+\lambda H$, $\lambda{=}0.1$ & 0.521 & 0.525 & 0.527 & 0.528 & 0.058 & 0.071 & 0.081 & 0.089\\
best-of-2 $+\lambda H$, $\lambda{=}0.01$ & 0.664 & 0.705 & 0.720 & 0.726 & 0.061 & 0.074 & 0.082 & 0.088\\
best-of-2 $+\lambda H$, $\lambda{=}0.03$ & 0.666 & 0.705 & 0.719 & 0.725 & 0.064 & 0.080 & 0.090 & 0.098\\
best-of-2 $+\lambda H$, $\lambda{=}0.1$ & 0.643 & 0.684 & 0.698 & 0.704 & 0.065 & 0.084 & 0.097 & 0.107\\
QuCoNet arch., single-attempt & 0.592 & 0.598 & 0.600 & 0.602 & 0.018 & 0.021 & 0.023 & 0.024\\
QuCoNet arch., best-of-2 & 0.699 & 0.733 & 0.745 & 0.750 & 0.014 & 0.017 & 0.019 & 0.021\\
QuCoNet arch., best-of-4 & 0.729 & 0.802 & 0.829 & 0.840 & 0.017 & 0.021 & 0.024 & 0.027\\
QuCoNet arch., best-of-8 & 0.707 & 0.820 & 0.871 & 0.896 & 0.017 & 0.022 & 0.025 & 0.028\\
QuCoNet arch., best-of-16 & 0.648 & 0.796 & 0.871 & 0.912 & 0.017 & 0.023 & 0.028 & 0.031\\
QuCoNet arch., best-of-32 & 0.564 & 0.735 & 0.835 & 0.894 & 0.019 & 0.027 & 0.033 & 0.038\\
QuCoNet arch., best-of-64 & 0.437 & 0.610 & 0.728 & 0.809 & 0.023 & 0.034 & 0.044 & 0.052
\end{tabular}
\end{ruledtabular}
\end{table}

\emph{What does the uniform policy diagnose?}  The untrained uniform walker,
evaluated on the same pools,
has mean first-arrival probability $\bar p=0.0088$ on both splits (the two
pools have the same graph-level mean under uniform routing).  On the held-out
split, classical sampling raises its accuracy to $0.026$, $0.043$, $0.060$,
$0.076$
at $q=3,5,7,9$ verified attempts, while fixed-depth amplification reaches $0.077$,
$0.204$, $0.367$, $0.544$; the corresponding training values differ by at
most $0.001$.  The $G_{\le q}$ values equal the fixed-depth values: every uniform
success probability in these pools lies so deep in the small-$p$ tail that the
optimal iteration count exceeds the largest tested $q$.  These numbers bound the trained systems
from two sides (main-text Table~I and its discussion): no trained system matches the uniform
walker's held-out accuracy after coherent inference, and the uniform walker
matches no trained system on the training questions.  Paired by question
pool at $q{=}9$, the held-out difference is $-0.127\pm0.032$ ($95\%$
confidence, eight pools) for Grover-trained QuCoNet minus the uniform
reference, and $+0.281\pm0.016$ ($32$ pools) against the strongest
classical control.  This uniform reference is also specific
to a small, solution-dense instance.  Under uniform routing the mean
first-arrival probability falls with graph size ($0.028$, $0.019$, $0.016$
on the random-regular held-out pools at $N=120$, $240$, $480$) and with
question difficulty ($0.102$ to $0.0044$ across graph distance $3$ to $7$
on the puzzle, the last at horizon $M=9$), and the amplified value falls
with it: the uniform walker's one-round held-out lead over the natively
trained walker on the random-regular family shrinks from $1.7\times$ at
$N=120$ ($0.217\pm0.025$ against $0.125\pm0.032$) through $1.4\times$ at
$N=240$ to statistical parity at $N=480$ ($0.129\pm0.017$ against
$0.113\pm0.034$), while the trained walker's lead over the classical
control grows from $1.8\times$ to $3.1\times$ over the same range, and at
graph distance $7$ the neural quantum AI model already matches the uniform
value ($0.187\pm0.017$ against $0.198$ after three rounds).  Nothing here
depends on the training set: the uniform walker's accuracy after coherent
inference comes from the uniform prior itself, a consequence of the small,
solution-dense benchmark
rather than a strategy available on hard instances.

\emph{Does coherent inference rescue a classically trained policy?}  The decisive test
of the objective grants every classically trained policy the quantum
inference circuit.  The capped walker ($c=0.25$), evaluated with one fixed-depth
round, reaches $0.706$ on training questions against the
Grover-trained $0.953$: only $45\%$ of its trained probability lies within
$[0.2,0.3]$, while $41\%$ has drifted above $0.3$, where the rotation
overshoots (its own design point; two fixed-depth rounds reduce it to $0.221$).
The neural capped policy fails the same way, drifting to a mean trained
$p=0.51$ where one round returns $0.46$.  The strongest hybrid of any
control is the QuCoNet-architecture best-of-$32$ walker plus one fixed-depth
round, at $0.919$, still below the Grover-trained value.  On held-out questions, controls trained with
large-$k$ objectives perform well after coherent inference at shallow depth
(best-of-$64$ plus fixed-depth inference: $0.132$, $0.205$, $0.255$, $0.299$ for
$q=3,5,7,9$) but fall
behind the models trained at larger $q$ from $q=7$ ($0.272$, $0.407$).  The
one-sided classical objectives neither constrain probabilities that drift
above the target nor sustain the gain at greater inference depth.  The
Grover-based objective supplies a gradient on both sides of the optimum.  A
two-sided classical objective with the same property is examined below.

\emph{Does the coin architecture alone prevent collapse?}  The same coins, trained on
$-\sum_i p_i$, undergo the same boundary concentration as CoNet.  For the question of the
main-text opening figure, this single-attempt-trained coin policy has $p=1.000$
and $\mathrm{IPR}=1.04$
[Fig.~\ref{fig:sm-flows}], against $p=0.250$ and
$\mathrm{IPR}=3.56$ under the one-round loss.  Thus unitarity of the parameter
matrix is insufficient; coherent inference must define the training target.

\emph{Can generic diversity regularization preserve probability for later
coherent inference?}  We add the Shannon-entropy bonus
$\lambda\sum_xH[\pi_x]$, where
$H[\pi_x]=-\sum_k\pi_x(k)\log\pi_x(k)$, to the single-attempt and best-of-two
objectives for $\lambda\in\{0.01,0.03,0.1\}$ at
$B\in\{8,32,128\}$, with $32$ question-pool seeds at $B=32$ and eight at
$B\in\{8,128\}$.  The
bonus either changes matched-cost accuracy by only a few points or degrades
it when uniform-routing pressure overwhelms reward.  It keeps some small-$p$
held-out mass alive but does not supply the question-wise stationary point of
Eq.~\eqref{eq:ridges}.

\emph{Can classical training import the interior target?}  The exact classical loss
$-\sum_i\min(p_i,c)$ with $c\in\{0.25,0.5,0.75\}$ stops rewarding a question
once $p_i$ exceeds the cap $c$.  At $c=0.25$ it is the strongest capped
control, and the stronger of the two controls tracked across training-set size in
Fig.~\ref{fig:sm-ladder}.  It demonstrates that classical training can use a
target supplied from outside the main-text monotonicity statement.  But
sampling raises a base probability $p=0.25$ only to $1-(0.75)^k$, while one
Grover round raises the interior optimum to unity.  The control is trained with
exact gradients and therefore has no Monte Carlo gradient-estimation noise;
the remaining cost difference occurs at inference.  For its per-question probabilities
$\{p_i\}$, let $k$ be the smallest number of independent verified attempts
whose mean best-of-$k$ accuracy matches the Grover-trained model.  At
$B=32$, $k\approx5\times10^{5}$ to match the $0.953$ training accuracy after
amplification ($4\times10^{5}$--$5\times10^{5}$ across the at-capacity sizes
$B=16$ and $32$; at smaller $B$ both models saturate and at larger $B$ the
lower coherent training accuracy is cheaper to match);
this tail-sensitive catch-up cost arises from the small-$p$ portion of the
learned distribution and is not implied by $C_{3}[\pstar(3)]$ alone.  Its
one-sided objective also has zero gradient for questions that drift above $c$.

To isolate that
asymmetry, we also test the symmetric loss
$\mathcal L_{\mathrm{2s}}=\sum_i(p_i-c)^2$ with $c=\pstar(q)$.  This removes
the one-sidedness and converges near the target: on the
QuCoNet architecture the trained probabilities lie within $c\pm0.05$
(fraction $0.91$ at $c=\pstar(3)$ and $1.00$ at the higher-order maxima), and
classical inference remains limited to the best-of-$k$ success of a
distribution at $c$.  Trained on the same coin architecture as the quantum
model, this control tracks the Grover-trained held-out sequence ($0.039$,
$0.157$, $0.303$, $0.479$ against $0.042$, $0.147$,
$0.272$, $0.407$ under fixed-depth inference), matching within seed spread
at $q\le5$ and sitting slightly above it at the larger $q$ (paired by
pool, the two-sided-minus-native held-out difference is $-0.001\pm0.006$,
$+0.007\pm0.007$, $+0.033\pm0.010$, $+0.061\pm0.011$ across
$q=3,5,7,9$), where the
Grover-based loss leaves probability at higher-order maxima that the
deepest fixed-depth protocol over-rotates; its held-out mean $p$ stays at
the uniform value.  The same holds for path diversity: over the eight shared
question pools, the accuracy-weighted training IPR of this control is
$1.62$, $3.25$, $6.14$, $9.31$ at $q=3,5,7,9$, against $1.62$, $2.83$,
$4.05$, $5.79$ for native Grover training, so a classical loss aimed at the
same interior target retains at least as many successful routes (the excess
at large $q$ mirrors the accuracy excess above: the two-sided loss pins
every question at the first maximum, whereas the native loss leaves some
weight at higher-order maxima).  The held-out
sequence, and the retained path diversity, are therefore set by the interior
target, not by the particular loss used to reach it.  Its fixed-depth training accuracy
after coherent inference is $0.952$, $1.000$, $1.000$, $1.000$, without the
higher-order peaks that reduce the native loss under the deepest fixed-depth
protocol (the $G_{\le q}$ rule absorbs
that difference).  On the softmax walker the same loss also lifts held-out
mean $p$ ($0.009\to0.028$ at $c=\pstar(3)$) and with it the fixed-depth sequence
($0.127$, $0.229$, $0.375$, $0.503$), but this is a transfer effect of the
shared per-node softmax rows on the small dense puzzle graph, not of the
loss: it is absent on the coin architecture, and on the random-regular
family the same control gains little and falls below the natively trained
walker at every size.  Both versions are oracle-assisted diagnostics rather
than deployed procedures: the constant $c=\pstar(q)$ is supplied by the
quantum analysis rather than derived from the inference dynamics the policy
will use.  The loss itself is classically estimable: the product of two
independent verifier verdicts, $(X-c)(Y-c)$ with
$X,Y\sim\mathrm{Bernoulli}(p)$, has expectation $(p-c)^2$, and pairing it
with the score function yields an unbiased gradient estimator, whose sample
complexity we do not evaluate here.
By contrast, the Grover target is derived from the deployed circuit: under
the fixed-depth protocol, the measured binary outcome is an unbiased sample
of $G_{q}(p)$ with no per-question input; realizing the adaptive
$G_{\le q}$ additionally requires the stopping choice $\widehat q(p;q)$,
which itself uses an estimate of $p$
(Sec.~\ref{sm:objective}).  The present
experiments nevertheless train with exact classically computed probabilities
and gradients; a hardware implementation would require a corresponding
stochastic-gradient estimator.

\emph{Optimization and selection.}  Tabular runs are evaluated every epoch
and neural runs every ten epochs; neural results report the final epoch.  Archived
tabular checkpoints are selected by the training value of the loss being
optimized for the classical best-of-$k$, capped, and two-sided objectives,
and by the highest raw single-attempt rate otherwise.  For single-attempt
objectives the two rules coincide.  For the Grover-based objective they do
not, because the raw rate is not monotone in the objective.  At moderate and
large training-set sizes, the interior optimum and IPR form a long plateau
after convergence and final-versus-best
selection is immaterial, but at the smallest $B/N$ ($B/N\lesssim0.1$) the raw
rate peaks during an early transient that overshoots the interior optimum, so the
archived checkpoint there predates convergence.  For this reason the
capacity scan of Sec.~\ref{sm:scaling} reads every run history at its
converged final epoch; reading the archived best-rate epoch instead
depresses the low-$B$ plateau into a spurious dip and shifts the capacity
thresholds by up to four percent.  A protocol-identical $200$-epoch classical rerun reproduces the
archived $120$-epoch baseline (single-attempt $0.257$ versus $0.252$ at $B=128$).
Some question pools at the largest $q$ converge to a deterministic boundary
solution rather than the lowest interior maximum.  We report all pools;
eliminating these
optimization failures is not part of the present claim.

The core exact-gradient tabular runs use Adam~\cite{kingma2015adam} for $200$ epochs with learning
rate $0.05$, softmax logits, and seed $42$ for each archived configuration;
reported tabular ensembles vary the question-pool seed, not this optimization
seed.  The neural
coin runs use $200$ epochs, learning rate $0.003$, gradient-norm clipping at
$1.0$, evaluation every ten epochs, and the architecture stated below.  These
settings, including objective and seed, are stored with each run in the
archived JSON configuration or history file.  The separate model-width sweep
uses at most $300$ epochs with early stopping after a minimum of $60$ epochs.

\section{A parameter-sharing neural quantum AI model}
\label{sm:neural}

The table-based model above uses one local $3\times3$ unitary coin operator per
node, parameterized by the six independent entries of a real symmetric
generator and hence $720$ real parameters in total.  A row is updated by every
question whose paths visit that node, so questions interact where their
reachable prefixes overlap.  We repeat the core measurements with the
table replaced by a single neural network.  A two-layer, four-head
prefix-conditioned transformer~\cite{vaswani2017attention}
($d_{\mathrm{model}}=32$ and $29{,}190$ parameters, with $d_{\mathrm{model}}$
denoting the hidden width) reads the question,
the target, and the move history of
a branch and emits the six independent entries of the
same $e^{iH}$ coin.  The circuit still uses one fresh record register per step.  Every question now
shares every parameter, and the coin conditions on the whole path prefix rather
than the current node.  The walk stays a record-controlled unitary,
block-diagonal in the which-path basis, so the classical-Markov training
structure and the objective theory of
Secs.~\ref{sm:objective}--\ref{sm:circuit} carry over unchanged.  Per-question
success is again evaluated exactly, here by a level-synchronous sweep of the
prefix tree, validated against independent enumeration to $4\times10^{-8}$.
Unless noted, $B=32$ on the puzzle and results use $32$ replicas; the same
replica index selects both the question pool and the neural-network
initialization.  All neural runs,
Grover-based and classical alike, use one implementation with
identical architecture, optimizer, learning rate, epoch count, and
checkpoint selection; the training loss is the only changed setting, so the
classical controls are not under-tuned relative to the Grover-trained runs.

\begin{figure}[t]
  \includegraphics[width=\linewidth]{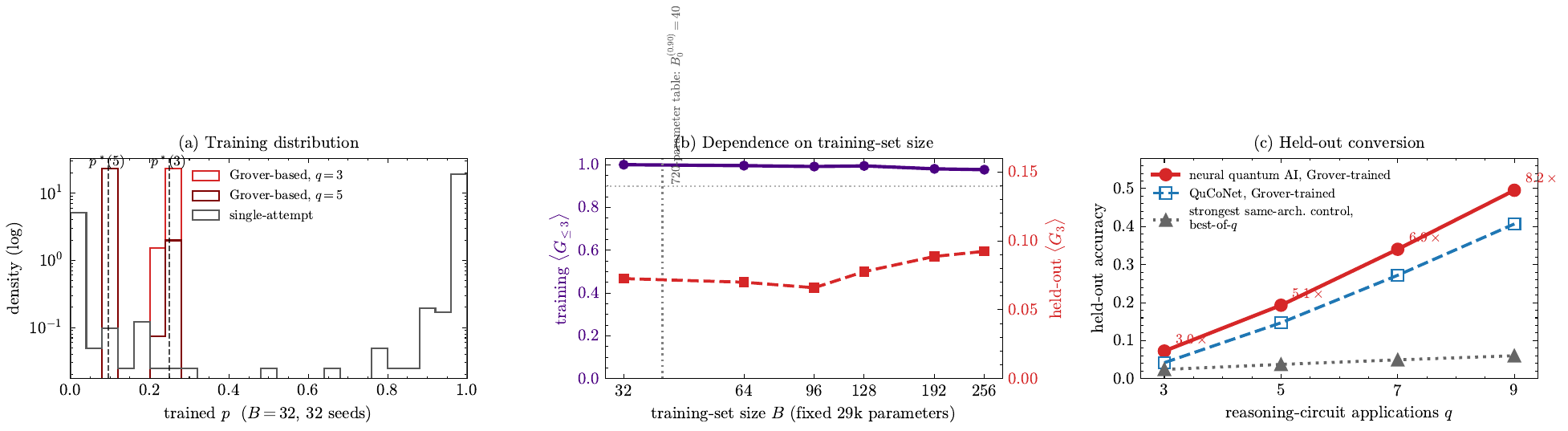}
  \caption{Grover-based training with a parameter-sharing neural quantum AI model
  (two-layer prefix-conditioned transformer, $29{,}190$ shared parameters;
  $32$ replicas unless noted).  (a)~At $B=32$, Grover-based training retains
  interior per-question probabilities near the predicted optima, whereas the
  single-attempt objective concentrates them near the boundaries.  (b)~At fixed model
  size, the training plateau persists through the tested range of $B$ while
  held-out accuracy increases; the dotted line marks the $720$-parameter
  table's capacity threshold at the $0.90$ level of the separate four-pool audit
  ($B_0^{(0.9)}=40$; the sixteen-pool $0.95$ definition of the main text
  gives $B_0(3)\approx33$).  (c)~At matched $q$, the neural quantum AI model exceeds both
  the per-node table and the strongest same-architecture classical control.}
  \label{fig:sm-neural}
\end{figure}

\emph{The interior optimum is set by the objective, not the parameterization.}
Table~\ref{tab:neural} collects the main matrix, and Fig.~\ref{fig:sm-neural}(a)
shows the trained distributions.  The Grover-based objective
places the neural policy near the predicted interior optimum ($q{=}3$: mean
initial success $p=0.2493\pm0.0036$, every question in the interior band
$0.02<p_i<0.98$, $\langle G_{\le3}\rangle=0.9995$),
whereas the same network under the single-attempt and best-of-two objectives
concentrates at the boundary (interior fraction $0.05$ and $0.23$).  At larger
$q$, the neural means lie systematically above the optima ($0.108$, $0.066$,
$0.045$ against $0.095$, $0.050$, $0.030$ for $q=5,7,9$), an offset of the
shared parameterization that costs little accuracy after coherent inference.
For the neural policy at $q\ge5$, the trained probabilities are therefore near,
rather than exactly at, the analytic optimum.  Boundary concentration is a
property of the classical target, reproduced on a policy whose memorization
structure, in which all parameters are shared, is the opposite of the table's.

\begin{table}[t]
\caption{Parameter-sharing neural quantum AI model at $B=32$ ($32$ replicas),
mean$\pm$standard deviation.
``Interior'' is the fraction of questions with $0.02<p_i<0.98$.
For the Grover-based rows, ``held-out'' is accuracy at the matched $q$ under
the fixed-depth protocol, using all $q$ circuit applications; the
optimal-stop rule adds at most
$0.003$, and the ratio is this held-out accuracy divided by the strongest
\emph{same-architecture} classical control's best-of-$q$ value; for the
control rows, ``held-out'' is single-attempt accuracy (their best-of-$q$
values appear in the text).}
\label{tab:neural}
\begin{ruledtabular}
\begin{tabular}{lcccc}
objective & initial success $p$ & interior & held-out & ratio\\
\midrule
Grover-based, $q=3$ & $0.249\pm0.004$ & $1.00$  & $0.073\pm0.017$ & $3.0\times$\\
Grover-based, $q=5$ & $0.108\pm0.010$ & $1.00$  & $0.193\pm0.025$ & $5.1\times$\\
Grover-based, $q=7$ & $0.066\pm0.007$ & $1.00$  & $0.340\pm0.030$ & $6.9\times$\\
Grover-based, $q=9$ & $0.045\pm0.004$ & $1.00$  & $0.496\pm0.043$ & $8.2\times$\\
single-attempt      & $0.776\pm0.081$ & $0.05$  & $0.008$         & ---\\
best-of-two         & $0.915\pm0.054$ & $0.23$  & $0.008$         & ---\\
capped $c{=}\pstar(3)$ & $0.512\pm0.040$ & $1.00$  & $0.009$         & ---
\end{tabular}
\end{ruledtabular}
\end{table}

\emph{Parameter sharing retains held-out performance; importing a target does
not import coherent inference.}
On held-out questions the neural policy trained at $q=3$ already exceeds the
strongest matched-cost control of its own architecture ($3.0\times$ at
three circuit applications)
[Fig.~\ref{fig:sm-neural}(c)], where the per-node table does not exceed the
matched classical control at $q{=}3$.  Against the strongest control of
\emph{any} parameterization (the
large-$k$ softmax family of Table~\ref{tab:controls}), the neural
sequence is $1.0$, $2.1$, $2.9$, $3.6$ times across $q=3,5,7,9$:
indistinguishable from parity at $q{=}3$ ($0.073\pm0.017$ versus
$0.070\pm0.026$) and ahead from $q{=}5$.  Among the neural controls
themselves, the single-attempt and best-of-two policies reach $0.022$ and $0.019$
at best-of-three and stay below the capped family at every $q$, so the
capped control is the strongest same-architecture comparator throughout.  It
keeps essentially every question interior (fraction $0.998$), but its mean
trained probability is $0.512$, not the supplied optimum $\pstar(3)=0.25$;
the one-sided cap therefore demonstrates interior retention rather than
faithful target placement.  The symmetric two-sided control above provides the
direct target-import diagnostic.  Granted
best-of-$q$ at each $q$, the capped controls ($c=\pstar(3)$ and
$c=\pstar(5)$, which agree to $0.002$) run
$0.024,0.038,0.049,0.060$ across $q=3,5,7,9$, while the Grover-trained policy runs
$0.073,0.193,0.340,0.496$: a held-out lead that widens monotonically from
$3.0\times$ to $8.2\times$ as $q$ increases [Fig.~\ref{fig:sm-neural}(c)], and
that beats the per-node table's corresponding values
($0.042,0.147,0.272,0.407$) at every $q$.  Granting the capped network quantum
inference instead, one fixed-depth
round raises its held-out accuracy to $0.063$, close to the
Grover-trained $0.073$ at $q{=}3$.  In these terminal-outcome-only sampling
controls, supplying the target does not supply coherent amplification, and
the gap grows with
difficulty and size: at graph distance $7$
(horizon $M=9$, eight replicas) the policy trained at $q=7$ reaches $0.187$ held-out,
$10.4\times$ the strongest same-architecture control, and on random
$3$-regular graphs at $N=480$
the policy trained at $q=5$ reaches $8.8\times$; the comparators for these two
claims are retrained on the harder task and the larger graphs,
respectively.

\emph{Capacity as a function of model size.}  Sweeping $B$ at fixed $29$k
parameters, the $q=3$ Grover-training plateau remains above the capacity
threshold throughout
the tested range [Fig.~\ref{fig:sm-neural}(b)]: $\langle G_{\le3}\rangle=0.9995,0.9952,0.9909,0.9939,0.9802,0.9762$
at $B=32,64,96,128,192,256$ (the last two are four replicas), still $0.976$ at
$B=256\approx8\,B_0(3)$ of the $720$-parameter table under the common $0.95$
definition.  The curve decreases gradually rather than showing the table
model's abrupt loss of capacity, and held-out accuracy ends higher at the
largest tested $B$ ($0.073\to0.092$) after a shallow dip through $B=96$, a gain
the per-node table cannot achieve.
For the width sweep we use a separate $0.9$ threshold.  Its $q=3$ capacity threshold moves
outward with parameter count: $B_0^{(0.9)}\approx114$ at
$d_{\mathrm{model}}{=}8$ ($5.9$k), while
$d_{\mathrm{model}}\ge16$ holds above $0.9$ through the largest pool tested
($B=256$, so the threshold is right-censored: it lies beyond the largest
training-set size tested).  This supports a qualitative outward shift of capacity
with model size, though it does not fit a size exponent.

\emph{The compiled neural policy remains unitary.}  The full-Hilbert-space
verification of Sec.~\ref{sm:circuit} repeats on compiled, prefix-controlled
coins from trained transformer instances: the amplitude state over all $K^{M}$
move sequences reproduces the absorbed prefix-tree probability to $8\times10^{-7}$
(\texttt{float32} network into a \texttt{float64} circuit), and one Grover round
through the reflection identity matches $\sin^{2}(3\vartheta)$ to $10^{-14}$ on
$q=3$ and $q=9$ Grover-trained instances alike.

The neural-coin held-out accuracies are small, so the $q{=}3$ ratios carry wide
relative uncertainty and the $B=192,256$ points use four replicas; the classical
controls are trained at $B=32$ and reused as the matched-cost comparator.  The
qualitative ordering (an interior optimum, collapse under one-sided
classical objectives, and a widening gap between coherent and classical
inference with difficulty and size) is stable across all replicas.

\section{Replication and scaling}
\label{sm:scaling}

\begin{figure}[t]
  \includegraphics[width=0.78\textwidth]{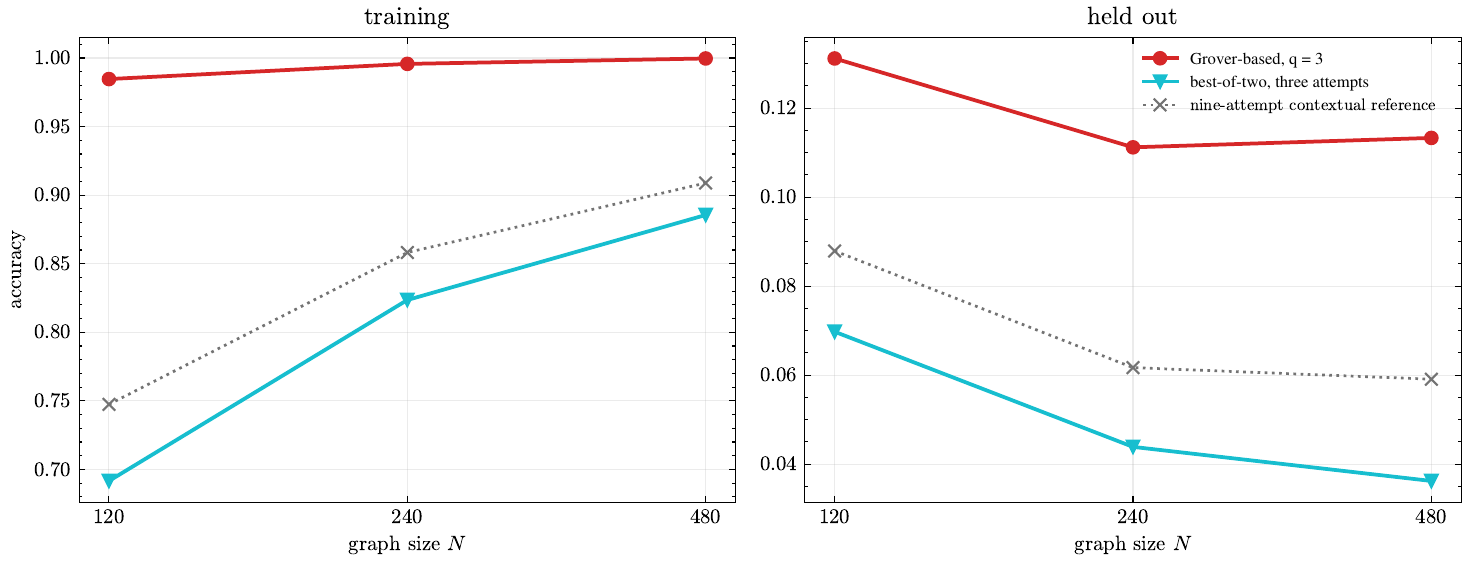}
  \caption{Problem-size control on random $3$-regular graphs at $B=32$.
  The one-round system (plotted under the $G_{\le3}$ optimal-stop rule; the
  fixed-depth values quoted in the text are up to $0.006$ lower) is compared
  with the exact best-of-two-trained
  classical model at three and nine samples.  The training-set difference narrows
  as $B/N$ falls; the held-out matched-cost difference persists.}
  \label{fig:sm-size}
\end{figure}

The random-regular family repeats three qualitative facts: single-attempt
training concentrates the unitary model at the boundary; one-round
Grover-based training retains an interior distribution; and training at larger
$q$ improves held-out accuracy after coherent inference.  Across the full $B$
grid, the $q=9$ system achieves $1.7$--$4.6$ times the best classical control at nine attempts in
Table~\ref{tab:controls} on the puzzle, and $2.6$--$5.4$ times on the
random-regular family, whose control set lacks the large-$k$ best-of-$k$
sequence.  On random-regular graphs, the held-out matched-cost lead remains at
$N=120,240,480$ [Fig.~\ref{fig:sm-size}].  The training-set gap narrows with
$N$ at fixed $B$, as expected when capacity pressure falls: the converged
coherent system holds training accuracy at or near unity at every size while
the classical control rises.  The held-out
ratio grows from approximately $1.8$ to $3.1$.

Two scaling relations clarify the mechanism.  First, a large-$B$ scan on the puzzle
(the $16$ extended pools of Sec.~\ref{sm:protocol}) uses a common $0.95$
cutoff on mean training accuracy after inference to define the capacity threshold
$B_0$.  Specifically, $B_0$ is the first downward crossing of
the $16$-pool mean curve, interpolated linearly in $\log B$; every run
history is read at its final epoch (Sec.~\ref{sm:controls}).  A small
fraction of cells still drifts at the training horizon (the mean accuracy
over the final quarter moves by more than $10^{-3}$ in $9\%$ of coherent and
$12\%$ of classical cells); reading a last-quarter average instead of the
final epoch moves both fitted exponents by less than $0.005$, so the
convention is immaterial at the reported precision.  At
$q=3$, all $16$ individual coherent curves reach the $0.95$ plateau.  Each bootstrap
draw resamples the $16$ question pools jointly for the coherent and classical
scans, preserving the within-pool comparison.  All $20{,}000$ draws
resolve every capacity threshold.  The measured capacity thresholds for
$q=3,5,7,9,11,13$ are
\begin{equation}
  B_0=33,64,107,168,274,450.
\end{equation}
The log--log fit against $q$ gives
$B_0\propto q^{1.739}$ ($R^2=0.976$), with central bootstrap intervals
$[1.729,1.755]$ at $68\%$ and $[1.718,1.771]$ at $95\%$.
A fit against $1/\pstar(q)$ has exponent $0.90$, and
$\pstar(q)B_0(q)$ ranges from $5.1$ to $8.2$.  The ordinary-least-squares
slope error is $0.14$, reflecting residual departure from a perfect power law
that resampling the question pools does not measure.  The exponent also inherits the
capacity-threshold definition: at threshold $0.90$ it is $2.067$, with a $95\%$ bootstrap
interval $[2.034,2.098]$, while at $0.85$ the largest-$q$ threshold lies beyond
the present $B$ range and is therefore right-censored.  A separate four-pool audit at the $0.90$ threshold
found that fixed-depth $G_q$ inference without adaptive stopping moves the
thresholds to
$40$, $68$, $125$, $203$, $317$, $467$, still monotone in $q$, with fitted
exponent $1.69$; the adaptive rule on the same four pools gives $2.04$,
consistent with the $16$-pool value above.  The protocol comparison is
threshold-dependent at the edges: at the $0.95$ level the fixed-depth mean
hovers at the criterion itself and does not produce a well-defined ladder
(the deepest protocol rotates early-maximum questions past their peaks,
Sec.~\ref{sm:objective}), while at a $0.70$ level the two protocols nearly
coincide over the range they resolve ($1.66$ against $1.71$ across
$q\le7$; the larger-$q$ thresholds lie beyond the tested $B$ range).
Classical best-of-$q$ success involves no stopping choice, so the
coherent--classical exponent difference of about one unit persists when
both sides are read under the unified fixed-depth protocol.  A
lower target asks less
success weight of each shared routing bottleneck, allowing a fixed policy to
hold more questions before frustration.

For the matched classical scan, we train the same $720$-parameter coin architecture
with the exact best-of-$q$ loss in Eq.~\eqref{eq:bestof-loss}, for
$q=3,5,7,9,11,13$, and evaluate each policy at the $q$ for which it was
trained.  All settings other than the outer response function are shared with
the coherent scan.  A position-Markov dynamic program computes the same
per-question probabilities as the full autoregressive state; a direct check
finds maximum probability error $3.7\times10^{-7}$, relative gradient error
below $3.6\times10^{-6}$, and probability disagreement $1.0\times10^{-6}$
after a matched $20$-epoch Adam trajectory.  At the same $0.95$ threshold,
the classical capacity thresholds are
\begin{equation}
  B_0^{\rm cl}=11.3,15.3,18.3,21.7,25.5,28.3.
\end{equation}
Their finite-range fit is $B_0^{\rm cl}\propto q^{0.624}$, with central
bootstrap intervals $[0.536,0.699]$ at $68\%$ and $[0.436,0.765]$ at $95\%$.
At threshold $0.90$ the exponent is $0.618$, with $95\%$ interval
$[0.573,0.712]$.  Thus the
classical plateau is real but confined to small training sets; a scan beginning at
$B=8$ without matched values of $q$ does not resolve it reliably.  At
matched $q$, the ratio $B_0/B_0^{\rm cl}$ grows from $2.91$ to $15.89$ across
the tested range.  The fitted exponents differ by $1.116$ (95\% bootstrap
interval $[0.979,1.306]$) under joint resampling of the same question pools.
This is a controlled finite-task separation, not an asymptotic
query-complexity statement.

\begin{figure}[t]
  \includegraphics[width=0.63\textwidth]{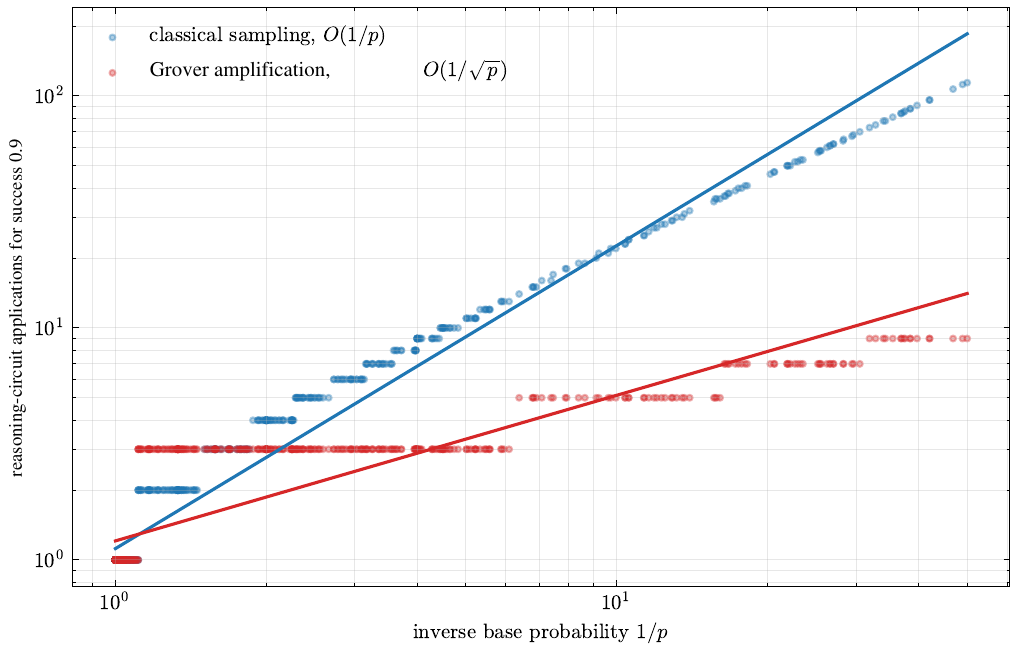}
  \caption{Finite-range check of the standard query scaling.  Here one query
  is one application of the reasoning circuit $\mathcal A$ or its inverse.
  The queries required to reach per-question success $0.9$ are plotted against
  inverse base probability.  Classical verified sampling approaches linear
  growth in $1/p$, whereas phase-matched amplitude amplification follows the
  square-root scale.  The final reflection uses a phase adjustment chosen to
  reach the target success without overshooting it by an integer Grover round.}
  \label{fig:sm-query}
\end{figure}

Second, for a fixed distribution of base probabilities, classical attempts
need $O(1/p)$ samples to reach constant success while amplitude amplification
with a phase-matched final reflection needs $O(1/\sqrt p)$
applications~\cite{brassard2002amplitude}.  The per-question query-count
diagnostic in Fig.~\ref{fig:sm-query} recovers fitted slopes $1.31$ and $0.63$ against
$1/p$ on the finite task.  The classical value exceeds unity because the exact
requirement $k(p)=\lceil\ln(1-\eta)/\ln(1-p)\rceil$, where $\eta$ is the
desired final success probability (here $\eta=0.9$), is only asymptotically
linear in $1/p$.  Ignoring the integer ceiling to display the local continuum
slope, one obtains
$\mathrm{d}\ln k/\mathrm{d}\ln(1/p)=p/[-(1-p)\ln(1-p)]$, which rises from $1.05$ at
$p=0.1$ to $1.44$ at $p=0.5$, so a single power law fitted across the resolved
window lands above one, while restricting the fit to the small-$p$ tail
returns an exponent within a few percent of unity.  Counting reasoning-circuit
applications directly as $q$, the quantum fit remains closer to its $1/2$ asymptote; integer
rounding over the resolved window raises the finite-range slope to $0.63$.
These are finite-range diagnostics of the standard
asymptotic maps, not a standalone end-to-end hardware speedup.  Only this
query-complexity diagnostic permits a phase-matched final reflection; all
policy accuracies and training objectives above use the fixed-phase map of
Eq.~\eqref{eq:Gleq-sm}.

\section{Scope, physical resources, and reproducibility}
\label{sm:limitations}

\emph{Computational boundary.}  Training, sweeps, and the full coherent circuit
are evaluated by exact classical calculation.  This is useful for testing the
mechanism because it removes sampling error and exposes the entire policy, but does
not establish a quantum hardware advantage.  A physical realization must pay
for reversible state preparation, $\mathcal A^\dagger$, the first-arrival
oracle, ancilla cleanup, and error correction.  For the neural experiment, our
full-state check validates the compiled prefix-controlled unitary; it does not
provide an efficient reversible implementation of the transformer.  Such an
implementation, or the cost of compiling its prefix-dependent coins, is an
additional resource not included in the query counts.

\emph{Coherence and memory.}  The circuit requires coherence across the full
horizon and fresh which-path storage.  Its logical record cost is
$M\lceil\log_2K\rceil$ qubits;
the present experiment embeds the complete state in $23$ qubits before oracle
ancillas.  Noise can turn the accepted/rejected rotation into classical mixing and erase
the benefit.  No robustness threshold is inferred from the noiseless data.

\emph{Reasoning scope.}  The task is Markovian, fixed horizon, and equipped
with an exact binary verifier.  The work does not yet model a language policy
with growing context, tool calls, self-modification, or a verifier that is
probabilistic or learned.  Which-path orthogonality means that the present
system does not require direct pairwise interference among semantic paths;
instead, the Grover reflection recombines their amplitudes through the
collective accepted and rejected sectors.  It establishes the smaller but
operationally precise primitive needed before richer models: coherent access
to the distribution of complete reasoning trajectories.

\emph{Reproduction.}  The training drivers, sweep launchers, analysis
scripts, and distilled probability caches are collected in the
paper's code repository; the principal analysis scripts
(under \texttt{docs/discussion/scripts/} and \texttt{notes/scripts/}) include:
\begin{itemize}
  \item \texttt{circuit\_verify.py} for
  Table~\ref{tab:circuit};
  \item \texttt{grover\_sweep\_analysis.py} for the interior optima and the
  grids at increasing $q$;
  \item \texttt{grover\_ipr\_analysis.py} for path support;
  \item \texttt{blind\_schedule\_analysis.py} for the fixed-depth
  evaluation, the capacity-threshold bootstrap and sensitivity, and the
  larger-$q$ IPR values;
  \item \texttt{control\_audit\_analysis.py} for the control audit
  (Table~\ref{tab:controls}), the untrained baseline, the cross-depth
  matrix (Table~\ref{tab:cross}), and the fixed-depth capacity thresholds;
  \item \texttt{classical\_capacity\_controlled.py},
  \texttt{capacity\_scaleup\_analysis.py}, and
  \texttt{plot\_capacity\_knee.py} for the matched $16$-pool capacity scan,
  bootstrap uncertainty from the shared question pools, threshold sensitivity,
  and the main-text comparison;
  \item \texttt{grover\_sizescan\_analysis.py} for
  Fig.~\ref{fig:sm-size};
  \item \texttt{twosided\_target\_control.py} and
  \texttt{twosided\_coin\_control.py} for the two-sided imported-target
  controls, \texttt{twosided\_ipr\_wave.py} for their path-diversity
  comparison, and \texttt{untrained\_scaling.py} for the untrained walker's
  size and difficulty dependence;
  \item \texttt{fixed\_depth\_capacity\_check.py} for the
  adaptive-versus-fixed-depth capacity protocol comparison, and
  \texttt{cost\_model\_sensitivity.py} for the verification-cost
  sensitivity;
  \item \texttt{robustness\_audits.py} for the pool--optimizer-seed
  cross-audit and the node-relabeling invariance test (relabeled pools
  built by \texttt{relabel\_qa.py});
  \item \texttt{amplification\_scaling.py} and
  \texttt{grover\_exchange\_rate.py} for query scaling.
\end{itemize}
The tracked file
\texttt{docs/discussion/scripts/\_sweep\_out/capacity\_scaleup\_16.json}
stores all pool-level capacity curves and the bootstrap summaries obtained by
jointly resampling the $16$ pools.  The other caches store the exact success probability of every
question under the reported trained models, so the tables and most figures
can be re-derived from the code repository alone.  Raw checkpoints and run records are
not included there and are required to regenerate the training sweeps.
All figures included in the manuscript are tracked artifacts so that the
LaTeX package compiles from a clean checkout.

\end{document}